\documentclass[prl,twocolumn,superscriptaddress]{revtex4-1}
\usepackage{}
\usepackage{amssymb}
\usepackage{amsfonts}
\usepackage{bbm}
\usepackage{mathrsfs}
\usepackage{graphics,graphicx,epsfig,bm,amsmath,amsthm,amssymb}
\usepackage{bm}
\usepackage{bbm}
\usepackage{longtable}
\usepackage{multirow}
\usepackage{array}
\usepackage{color}
\usepackage[usenames,dvipsnames]{xcolor}
\usepackage{float}
\usepackage[a4paper,colorlinks=true,
linkcolor=blue,citecolor=blue,
pdfauthor={ },
pdftitle={ },
pdfsubject={ },
pdfkeywords={ }]{hyperref}

\begin{document}

\title{Experimental time-optimal universal control of spin qubits in solids}

%\author{Jianpei Geng$^{1}$$^*$, Yang Wu$^{1}$$\thanks{These authors contributed equally to this work.}$, Xiaoting Wang$^{3,4}$, Kebiao Xu$^{1}$, Fazhan Shi$^{1,2}$,  Yijin Xie$^{1}$, Xing Rong$^{1,2}$\thanks{e-mail: xrong@ustc.edu.cn}, Jiangfeng Du$^{1,2}$\thanks{e-mail: djf@ustc.edu.cn}}
%\maketitle
%\begin{affiliations}
\author{Jianpei Geng}
\affiliation{CAS Key Laboratory of Microscale Magnetic Resonance and Department of Modern Physics , University of Science and Technology of China, Hefei 230026, China}
\author{Yang Wu}
\affiliation{CAS Key Laboratory of Microscale Magnetic Resonance and Department of Modern Physics , University of Science and Technology of China, Hefei 230026, China}
%\thanks{These authors contributed equally to this work.}
\author{Xiaoting Wang}
\affiliation{Hearne Institute for Theoretical Physics, Department of Physics and Astronomy, Louisiana State University, Baton Rouge, Louisiana 70803, USA}
\affiliation{Research Laboratory of Electronics, Massachusetts Institute of Technology, Cambridge, Massachusetts 02139, USA}
\author{Kebiao Xu}
\affiliation{CAS Key Laboratory of Microscale Magnetic Resonance and Department of Modern Physics , University of Science and Technology of China, Hefei 230026, China}
\author{Fazhan Shi}
\affiliation{CAS Key Laboratory of Microscale Magnetic Resonance and Department of Modern Physics , University of Science and Technology of China, Hefei 230026, China}
\affiliation{Synergetic Innovation Center of Quantum Information and Quantum Physics, University of Science and Technology of China, Hefei 230026, China}
\author{Yijin Xie}
\affiliation{CAS Key Laboratory of Microscale Magnetic Resonance and Department of Modern Physics , University of Science and Technology of China, Hefei 230026, China}
\author{Xing Rong}
\email{xrong@ustc.edu.cn}
\affiliation{CAS Key Laboratory of Microscale Magnetic Resonance and Department of Modern Physics , University of Science and Technology of China, Hefei 230026, China}
\affiliation{Synergetic Innovation Center of Quantum Information and Quantum Physics, University of Science and Technology of China, Hefei 230026, China}
\author{Jiangfeng Du}
\email{djf@ustc.edu.cn}
\affiliation{CAS Key Laboratory of Microscale Magnetic Resonance and Department of Modern Physics , University of Science and Technology of China, Hefei 230026, China}
\affiliation{Synergetic Innovation Center of Quantum Information and Quantum Physics, University of Science and Technology of China, Hefei 230026, China}

%\end{affiliations}

%\date{\today}

\begin{abstract}
Quantum control of systems plays an important role in modern science and technology.
The ultimate goal of quantum control is to achieve high fidelity universal control in a time-optimal way.
Although high fidelity universal control has been reported in various quantum systems, experimental implementation of time-optimal universal control remains elusive.
Here we report the experimental realization of time-optimal universal control of spin qubits in diamond.
By generalizing a recent method for solving quantum brachistochrone equations [X. Wang \textit{et al.}, Phys. Rev. Lett. \textbf{114}, 170501 (2015)], we obtained accurate minimum time protocols for multiple qubits with fixed qubit interactions and a constrained control field.
Single- and two-qubit time-optimal gates are experimentally implemented with fidelities of $99\%$ obtained via quantum process tomography.
Our work provides a time-optimal route to achieve accurate quantum control and unlocks new capabilities for the emerging field of time-optimal control in general quantum systems.
\end{abstract}

%\pacs{Valid PACS appear here}

\maketitle
Time-optimal control (TOC), including the famous examples of the brachistochrone problem \cite{IEEEControlSystMag_Sussmann} and the Zermelo navigation problem \cite{ZAngewMathMech_Zermelo}, has been widely investigated for over three centuries.
TOC of quantum systems has recently attracted great interest due to the rapid development of quantum information processing and quantum metrology.
Because the ever-present noise from the environment degrades quantum states or operations over time, generating the fastest possible evolution by TOC becomes a preferable choice for realizing precise quantum control in the presence of noise.
%Because of the degradation of quantum states or operations over time due to the ever-present noise, generating fastest possible evolution by TOC becomes a preferable choice for realizing precise quantum control.
%TOC is also shown to be related to gate complexity in quantum computation, thus provides an alternative method to understand and analyze complexity of a quantum algorithm \cite{Science_Nielsen, PRA_Koike}.
%Particularly in quantum computation, it has been shown that the time-optimal solution of a unitary operation is related to the gate complexity, which provides an alternative method to understand and analyze complexity of a quantum algorithm\cite{Science_Nielsen, PRA_Koike}.
To obtain accurate TOC protocols is difficult because both the fidelity and time should be optimized.
Analytical methods utilizing the Pontryagin maximum principle or the geometry of the unitary group are applicable only to specific problems and constraints \cite{PRA_Glaser_2001, PRA_Brockett, PRA_Yuan_2005, PRA_Fisher_2009, PRA_Boozer, PRL_Hegerfeldt, PRA_Garon, PRA_Yuan_2015}.
Recently, the quantum brachistochrone equation (QBE) has been proposed to provide a general framework for finding time-optimal state evolutions or unitary operations \cite{PRL_Carlini, PRA_Carlini, PhDThesis_Okudaira, PRL_Rezakhani, JPA_Carlini_2011, PRA_Carlini_2012, JPA_Carlini_2013, PRL_Mohseni}.
%In view of this, a significant development is the quantum brachistochrone equation (QBE) which provides a general framework for finding time-optimal state evolutions or unitary operations \cite{PRL_Carlini, PRA_Carlini, PhDThesis_Okudaira, PRL_Rezakhani, JPA_Carlini_2011, PRA_Carlini_2012, JPA_Carlini_2013, PRL_Mohseni}.
The QBE has been applied to some cases where analytic solutions exist \cite{JPA_Carlini_2011, PRA_Carlini_2012, JPA_Carlini_2013}.
For problems where the QBE cannot be analytically solved, an effective numerical method has been developed \cite{PRL_Mohseni}.
%Recently, an effective numerical method for solving the QBE has been developed \cite{PRL_Mohseni}, which unlocks the applications of QBE to problems where the QBE cannot be analytically solved.
%For some special cases analytic solutions to QBE can be obtained \cite{JPA_Carlini_2011, PRA_Carlini_2012, JPA_Carlini_2013}. For general TOC of multiple qubits an effective numerical method has been recently developed to solve the QBE \cite{PRL_Mohseni}.
%the QBE has to be numerically solved.
%Some cases such as time-optimal coherence transfer or unitary operations in an Ising chain have been discussed by analytically solving the QBE \cite{JPA_Carlini_2011, PRA_Carlini_2012, JPA_Carlini_2013}.
%Fortunately an effective numerical method for solving the QBE has been recently developed \cite{PRL_Mohseni}.
%These theoretical progresses have made TOC of multiple qubits possible.
The relationship between TOC and gate complexity has also been explored \cite{Science_Nielsen, PRA_Koike}.
Experimental TOC has been implemented only in single-qubit systems \cite{NaturePhys_Morsch,PRL_Sugny, PRB_Gershoni},
%Experimental progress has been made only in preparing a desired state \cite{NaturePhys_Morsch,PRL_Sugny} or unitary operations \cite{PRB_Gershoni} in a two-level system.
while experimental time-optimal universal control, which requires universal single-qubit gates as well as a non-trivial two-qubit gate, has not been reported.
%Experimental time-optimal universal control, which requires universal single-qubit gates as well as a non-trivial two-qubit gate (e.g. a controlled-not gate), has not been reported.

Here, we demonstrate the first experimental time-optimal universal control of a two-qubit system, which consists of an electron spin and a nuclear spin of a nitrogen-vacancy (NV) center in diamond.
% We develop the method for solving the QBE equation to obtain the TOC with time invariable interactions and constrained control filed, which makes the TOC of multiple qubits experimentally feasible.
High-fidelity single- and two-qubit gates are realized with fidelities of $99\%$ obtained via quantum process tomography.
Our results show that TOC provides a novel route to achieve precise universal quantum control.
The approach to realize time-optimal control of multiple qubits can be applied to other quantum systems.
%The solution to QTOC can find applications in various areas. For example, a short time corresponds to an efficient strategy in adiabatic quantum computation\cite{PRL_Rezakhani}. Besides,

\begin{figure}\centering
\includegraphics[width=0.9\columnwidth]{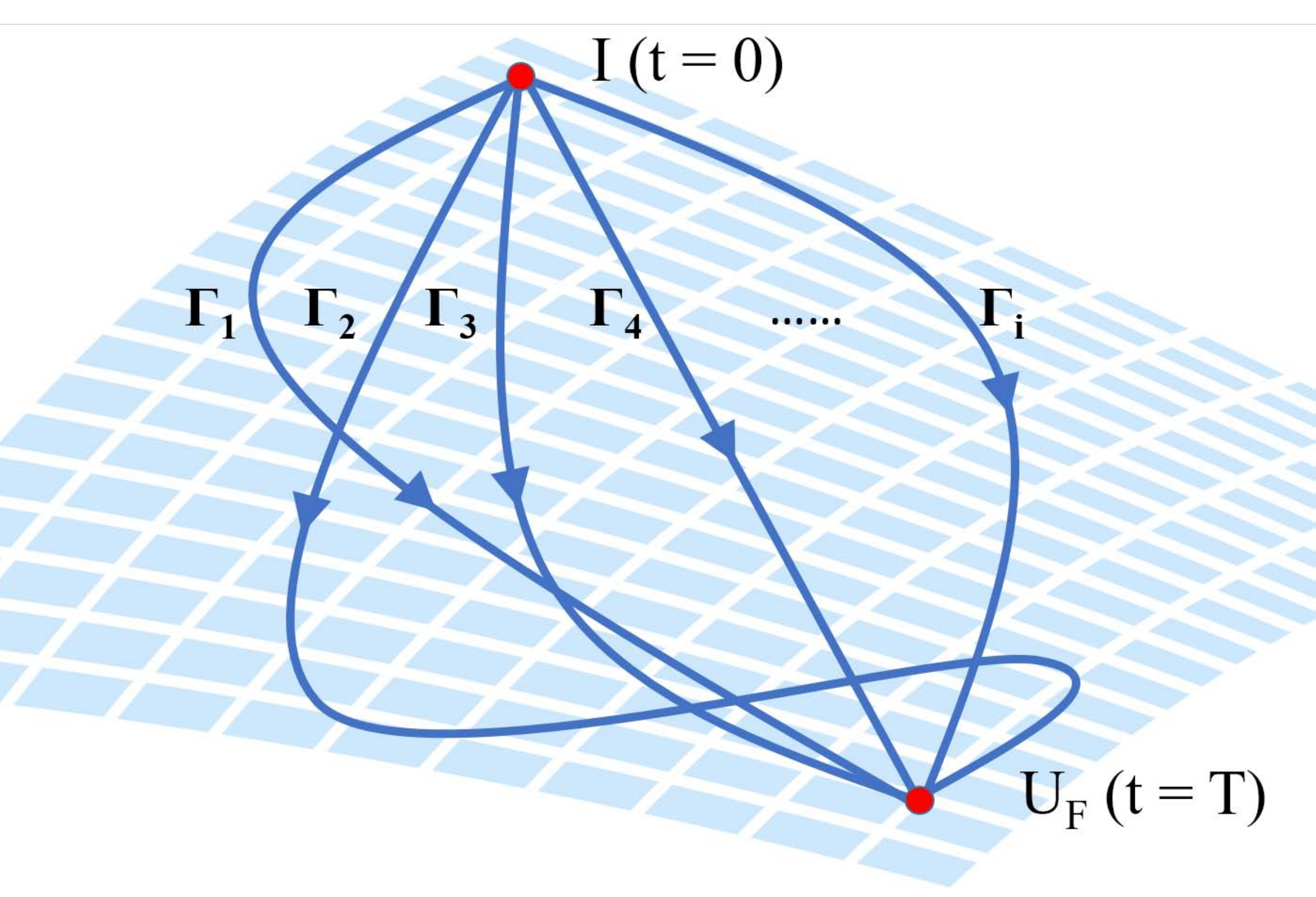}
\caption{(color online). Schematic representation of quantum TOC. Blue lines represent paths of quantum evolution in the SU($2^n$) operator space, where $n$ stands for the number of qubits. To realize a target evolution operator $U_\textrm{F}$ at $t = T$ starting from the identity operator $I$ at $t = 0$, there are several choices of evolution path $\Gamma_i~(i = 1,2\ldots)$. The goal of TOC is to figure out which evolution costs the minimum time $T$.
 }
    \label{Fig1}
\end{figure}

%\section{Results}
%\subsection{Quantum brachistochrone equation.}
%The aim of TOC is to realize the target evolution $U_F$(labeled by red circle) from the identity operator(labeled by black cirlce) with the minimal time.
As shown in Fig.~\ref{Fig1}, the quantum system is driven by the Hamiltonian $H(t) $, which is described by the Schr\"{o}dinger equation $\dot{U} = - i H(t) U$, with boundary conditions $U(0) = I $ and $U(T) = U_F$ (we set $\hbar = 1$).
Different Hamiltonians $H(t)$ make the evolutions of the system follow different paths (labeled by $\Gamma_i$) to the same unitary operation $U_F$.
The path with the minimal time cost can be obtained by solving the QBE \cite{PRA_Carlini} together with the Schr\"{o}dinger equation. The QBE is written as
\begin{equation}
\label{QBE}
\dot{F}=-i[H, F],
\end{equation}
where $F=\partial L_C/\partial H$ and $L_C=\sum_j\lambda_jf_j(H)$, with $\lambda_j$ the Lagrange multiplier.
One physically relevant constraint is the finite energy bandwidth described as $f_0(H)\equiv[{\rm Tr}(H^2)-E^2]/2=0$, where $E$ is a constant. Reference~\onlinecite{PRL_Mohseni} provides a method to obtain the accurate minimum-time protocol by solving the QBE.

In realistic physical systems, part of the Hamiltonian $H$ is usually time independent (e.g., fixed couplings between spin qubits), and the reasonable constraint for the energy is actually for the time variable part (e.g., the shaped microwave pulse with bounded power).
These have been recently recognized and investigated as the quantum Zermelo navigation problem \cite{PRL_Meier,PRA_Stepney}.
The original QBE is not able to provide a solution to this problem directly.
Here, we rewrite the Hamiltonian $H(t)$ as $H = H_0 + H_c(t) $, where the drift Hamiltonian $H_0$ stands for the time invariable part and $H_c(t)$ stands for the control Hamiltonian.
The drift Hamiltonian $H_0$ can be the fixed spin couplings or nonzero constant external magnetic field.
The control Hamiltonian $H_c(t)$ can be a controllable external magnetic field or adjustable couplings between qubits.
The constraint of the finite energy bandwidth is modified to $f_0(H_c)=0$.
Then, the TOC of multiple qubits, which is experimentally feasible,  can be obtained by solving the QBE with the mentioned improvements (see Section \uppercase\expandafter{\romannumeral2} in Supplementary Material).
%Our theoretical approach works well for finding the time-optimal way to steer the general quantum systems.
%, of which the qubits' interactions can be controllable  or uncontrollable.
This method can be taken as the generalization of the method in Ref.~\onlinecite{PRL_Mohseni}, which is the case of $H_0 = 0$.

%\subsection{NV center in diamond.}
We experimentally demonstrate TOC of single- and two-qubit on an NV center in diamond.
The NV center is composed of an electron spin and a nitrogen nuclear spin.
A static magnetic field of about 500 G is applied along the NV symmetry axis ([1 1 1] crystal axis) and removes the degeneracy between the $|m_S=+1\rangle$ and $|m_S=-1\rangle$ electron spin states.
Under such a magnetic field, the spin state of the NV center is effectively polarized to $|m_S=0, m_I=+1\rangle$ when a 532 nm laser pulse is applied \cite{PRL_Wrachtrup}.
Microwave pulses driving the electron spin transition $|m_S=0\rangle$ to $|m_S=-1\rangle$ and radio-frequency pulses driving the nuclear spin transition $|m_I=+1\rangle$ to $|m_I=0\rangle$ are utilized to manipulate the spin states.
The $|m_S=+1\rangle$ electron spin level and $|m_I=-1\rangle$ nuclear spin level remain idle due to large detuning.
TOC is demonstrated on the two-qubit system composed by $|m_S=0, m_I=+1\rangle$, $|m_S=-1, m_I=+1\rangle$, $|m_S=0, m_I=0\rangle$, and $|m_S=-1, m_I=0\rangle$ without considering the other spin levels (see Section \uppercase\expandafter{\romannumeral1} and Fig. S1 in Supplementary Material).

The experiment was implemented on an NV center in $[100]$ face bulk diamond.
The nitrogen concentration in the diamond was less than 5 ppb and the abundance of $^{13}$C was at the natural level of 1.1\%.
The NV center was optically addressed by a home-built confocal microscope. Spin-state initialization and detection of the NV center were realized with a 532 nm green laser controlled by an acousto-optic modulator (ISOMET, power leakage ratio $\sim$1/1000).
To preserve the NV center's longitudinal relaxation time from laser leakage effects, the laser beam was passed twice through the acousto-optic modulator before going through an oil objective (Olympus, PLAPON 60*O, NA 1.42).
The phonon sideband fluorescence (wavelength 650$\sim$800 nm) went through the same oil objective and was collected by an avalanche photodiode (Perkin Elmer, SPCM-AQRH-14) with a counter card.
A solid immersion lens was created around the NV center to increase the fluorescence collection efficiency.
The magnetic field was provided by a permanent magnet and aligned by monitoring the variation of fluorescence counts.
The spin states of the NV center were manipulated with microwave and radio-frequency pulses.
The microwave and radio-frequency pulses were generated by an arbitrary waveform generator (Keysight M8190A), amplified individually with power amplifiers (Mini Circuits ZHL-30W-252-S+ for microwave pulses and LZY-22+ for radio-frequency pulses), and combined with a diplexer (Marki DPX-1).
An ultra-broadband coplanar waveguide with $15~$GHz bandwidth was designed and fabricated to feed the microwave and radio-frequency pulses.

%\subsection{Single-qubit time-optimal control.}
Universal control of a single qubit requires the ability to realize rotations around two different axes of the Bloch sphere.
The evolution operator is denoted with $R (\hat {\mathbf n},\theta)$, corresponding to a rotation of angle $\theta$ around axis $\mathbf{\hat n}=\mathbf{\hat x}\sin\gamma\cos\varphi+\mathbf{\hat y}\sin\gamma\sin\varphi+\mathbf{\hat z}\cos\gamma$.
The method to realize TOC gates, which rotate the quantum states along two different axes, is detailed in Section \uppercase\expandafter{\romannumeral2} in Supplementary Material.
We take a target unitary transformation $R (\hat {\mathbf z} ,\theta)$ on the electron spin qubit as an example.
In the rotating frame, $H_0=2\pi\delta S_z$, $H_c(t)=2\pi\nu_1[\cos\phi(t)S_x+\sin\phi(t)S_y]$,
where $S_x$, $S_y$, and $S_z$ are effective spin operators of the electron spin qubit, $\delta $ is the detuning term, $\nu_1>0$ stands for the amplitude of the microwave  pulse, and $\phi(t)$ is the phase of microwave pulse.
The control Hamiltonian $H_c$ satisfies two constraints, which are $f_0(H_c)\equiv[{\rm Tr} (H_c^2)-2\pi^2\nu_1^2]/2=0$ and $f_1(H_c)\equiv {\rm Tr}(H_cS_z)=0$.
The solution to the QBE is $\phi(t)=2\pi\eta t+\phi(0)$, where $\eta$ is a constant.
Then the detailed parameters of the control Hamiltonian [e.g., $\eta$ and $\phi (t)$] and the minimum evolution time $T$ can be obtained by further solving the Schr\"{o}dinger equation.
By following the procedure described above, we can derive the explicit
analytical solutions to the TOC for realizing  $R (\hat {\mathbf z} ,\theta)$.
Without loss of generality, we present the analytical solution when $\delta\geq0$ and $\theta\in[0, 2\pi)$.
If $\theta <  \pi (1 + \sqrt{3} \delta / \nu_1)$, the minimum evolution time $T=[\delta(\theta/2\pi-1)+\sqrt{\nu_1^2+\delta^2-\nu_1^2(\theta/2\pi-1)^2}]/(\nu_1^2+\delta^2)$; otherwise, the minimum evolution time becomes $T=[\delta\theta/2\pi+\sqrt{\nu_1^2+\delta^2-\nu_1^2(\theta/2\pi)^2}]/(\nu_1^2+\delta^2)$.
The minimum evolution time $T$ versus $\theta$ and $\delta/\nu_1$ is shown in Supplementary Fig. S2.
The case when $\delta\neq0$ is of importance to those systems where it is challenging to adjust the detuning, such as the singlet-triplet spin qubit in a double-quantum-dot system \cite{NaturePhys_Foletti, NatureCommun_XinWang}.
When $\delta = 0$, our result reduces to that in Ref.~\onlinecite{PRA_Boozer}.
%In this case, the time for TOC is compared to that with a typical Euler rotation.
%As expected, the time for TOC is shorter than for the Euler rotation, which is verified by both theoretical and experimental results (see Section \uppercase\expandafter{\romannumeral2}, Fig. S3, and Fig. S4 in Supplementary Material).
%Both theoretical and experimental results show that the time is shorter with TOC than with Euler rotation.

\iffalse
\begin{figure}\centering
\includegraphics[width=0.9\columnwidth]{Fig2.pdf}
\caption{(color online). Experimental implementation of single-qubit TOC for target evolution operators (a)(b) $R (\hat {\mathbf z} ,\pi/2)$ and (c)(d) $R (\hat {\mathbf x} ,\pi/2)$. (a)(c) State population of $|m_S=0\rangle$ during the time-optimal implementation of (a) $R (\hat {\mathbf z} ,\pi/2)$ and (c) $R (\hat {\mathbf x} ,\pi/2)$ with initial state $|m_S=0\rangle$. Inset: Energy level diagrams showing schematically the populations corresponding to initial and final states. (b)(d) The left panels show the time-optimal evolution paths, and the right panels are experimental results of quantum process tomography. Form the results of quantum process tomography, an average gate fidelity of 1.00(1) and 0.99(1) for (b) $R (\hat {\mathbf z} ,\pi/2)$ and (d) $R (\hat {\mathbf x} ,\pi/2)$ can be obtained, respectively.
 }
    \label{Fig2}
\end{figure}
\fi

\begin{figure}\centering
\includegraphics[width=0.9\columnwidth]{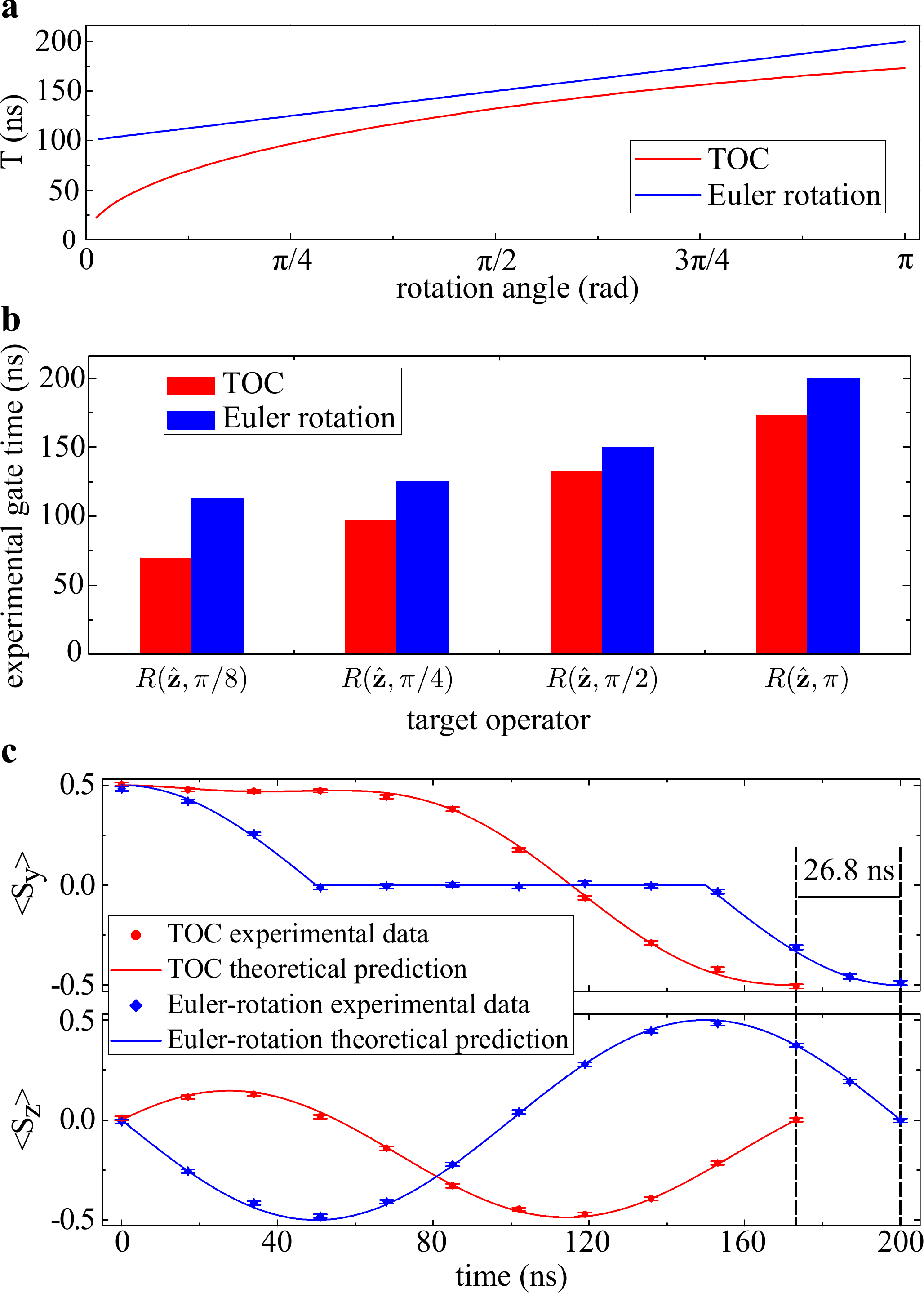}
\caption{(color online). Comparison on time costs for target gate operator $R (\hat {\mathbf z} ,\theta)$ between the derived TOC and the Euler rotations. The parameters are set to be $\delta=0$ and $\nu_1=5$~MHz. (a) Theoretical comparison on time with $\theta\in(0, \pi]$. (b) Comparison of experimental gate time for $\theta=\pi/8, \pi/4, \pi/2,$ and $\pi$. The gate time for TOC is considerably shorter than that for Euler rotation. (c) State evolutions during $R (\hat {\mathbf z} ,\pi)$ with TOC and Euler rotation. The initial state is $(|m_s=0\rangle+i|m_s=-1\rangle)/\sqrt{2}$. The gate time of TOC is 26.8 ns shorter than that of Euler rotation.
 }
    \label{Fig2}
\end{figure}

The realization of a target $R (\hat {\mathbf z} ,\theta), \theta\in(0,\pi]$ when $\delta=0$  is taken as an example to compare the time cost between the derived TOC and a non-optimized evolution path with Euler rotation: $R (\hat {\mathbf z} ,\theta)=R (\hat {\mathbf x} ,\pi/2)R (\hat {\mathbf y} ,\theta)R (-\hat {\mathbf x} ,\pi/2)$.
%The comparison on time between the TOC and Euler rotation is shown in Fig.~\ref{Comparison with Euler rotation}.
The experimental amplitude of control field is set to be $\nu_1=5$~MHz.
Theoretical comparison of  the time cost for gate operations between TOC and Euler rotation is shown in Fig.~\ref{Fig2}a.
It is clear that the time cost with TOC is considerably shorter than that with Euler rotation for all the rotation angles.
We experimentally implement the target gate operators $R (\hat {\mathbf z} ,\pi/8)$, $R (\hat {\mathbf z} ,\pi/4)$, $R (\hat {\mathbf z} ,\pi/2)$, and $R (\hat {\mathbf z} ,\pi)$ with both methods.
Figure \ref{Fig2}b shows the comparison of the experimental gate time.
The time durations for gate operations with TOC are 69.6, 96.8, 132.3, and 173.2 ns, which are 42.9, 28.1, 17.7, and 26.8 ns shorter than those with Euler rotation, respectively.
Figure \ref{Fig2}c shows the state evolution during $R (\hat {\mathbf z} ,\pi)$.
%The target operator $R (\hat {\mathbf z} ,\pi)$ is taken as an example.
The initial state is prepared to $(|m_s=0\rangle+i|m_s=-1\rangle)/\sqrt{2}$.
We performed measurements of $\langle S_y\rangle$ and $\langle S_z\rangle$ on the states during the evolution.
As shown in Fig.~\ref{Fig2}c, the target evolution of $R (\hat {\mathbf z} ,\pi)$ is realized at 173.2~ns with TOC and at 200~ns with Euler rotation.
All the gate fidelities \cite{PLA_Nielsen}  are measured to be above 0.99 via quantum process tomography \cite{Nielsen}.

The case when $\delta\neq0$ has also been experimentally implemented.
Both $R (\hat {\mathbf z} ,\theta)$ and $R (\hat {\mathbf x} ,\theta)$ with various values of $\theta$ have been demonstrated.
Furthermore, time-optimal universal single-qubit control with other constraints on $H_c$ is also experimentally demonstrated.
The implementations are characterized utilizing quantum process tomography (see Section \uppercase\expandafter{\romannumeral3} in Supplementary Material).
%A physical process can be represented by a completely positive map $\varepsilon$ on an arbitrary input state $\rho$: $\varepsilon(\rho)=\sum_{mn}\chi_{mn}A_m\rho A_n^\dag$, with operators ${A_n}$ forming a basis and $\chi$ being the process matrix.
%The process matrix $\chi$ determined by quantum process tomography contains all the information of the physical process.
The experimental results for the cases are presented in Section \uppercase\expandafter{\romannumeral2}, Fig. S3, Fig. S4, and TABLE \uppercase\expandafter{\romannumeral1} of the Supplementary Material.
Our results show the universality of our approach to perform time-optimal universal control for a single qubit.
Universal control of qubits also requires a non-trivial two-qubit gate \cite{JAJones}. In our experiment, we demonstrate a controlled-U gate with
\begin{equation}
U_c = \left(
      \begin{array}{cccc}
        1& 0 & 0 & 0 \\
        0 & 0 & 0 & 1 \\
        0 & 0 & 1 & 0 \\
        0 & -1 & 0 & 0 \\
      \end{array}
    \right),
\end{equation}
which is also a non-trivial two-qubit gate \cite{JAJones}.
In our experiment, we have demonstrated this two-qubit gate in a time-optimal way with the system consisting of  the electron and nuclear spins.
Electron (nuclear) spin states $|m_S=0\rangle$ and $|m_S=-1\rangle$ ($|m_I=+1\rangle$ and $|m_I=0\rangle$) are encoded as the electron (nuclear) spin-qubit.
The quantum state of the two-qubit system is denoted as $|m_S, m_I\rangle$, with corresponding population denoted as $P_{m_S, m_I}$ hereafter.
The drift Hamiltonian, $H_0=2\pi AS_zI_z$, is the hyperfine coupling between the spins, where $I_z$ is the effective spin operator of the nuclear spin qubit and the hyperfine coupling strength is $A=-2.16~$MHz.
We consider a model in which only controls with bounded strength on the electron spin are applied, while the control Hamiltonian takes the form $H_c(t)=2\pi\nu_1[\cos\phi(t)S_x+\sin\phi(t)S_y]$.
The strength of the control field $\nu_1$  is set to $2.5~$MHz.
The constraints on the control Hamiltonian can be described by $f_0(H_c)=0$ and $f_k(H_c)\equiv Tr(H_cB_k)=0$, where $\{B_k\}=\{I_x, I_y, I_z, S_xI_x, S_xI_y, S_xI_z, S_yI_x, S_yI_y, S_yI_z, S_z, S_zI_x, \\S_zI_y, S_zI_z\}$.
The target evolution operator is a controlled unitary gate which flips the electron spin qubit iff the nuclear spin qubit is in the state $|m_I=0\rangle$.
The time-optimal control Hamiltonian is obtained by numerically solving the QBE together with the Schr\"{o}dinger equation (see Section \uppercase\expandafter{\romannumeral2} in Supplementary Material).
If the dephasing effect and the imperfection of the control field are taken into account \cite{NatCommun_Du}, the theoretical fidelity of $U_c$ is estimated to be $0.9933$. The detailed experimental pulse for time-optimal control and the fidelity estimation are included in Section \uppercase\expandafter{\romannumeral2} and Fig. S5 in Supplementary Material.
The time duration of the controlled-U gate with TOC is 446 ns.
A conventional method to implement the controlled-U gate with the constraint control field is to apply a selective pulse \cite{PRA_Dorai, JMR_Mahesh}.
With $\nu_1=2.5$~MHz (the same as that in TOC), the time duration to implement the controlled-U gate with a selective pulse is 612.4~ns (see Section \uppercase\expandafter{\romannumeral2} in Supplementary Material), which is more than 160 ns longer than that with TOC.
%The target evolution operator is $U_{\textrm{F}, 2\textrm{-q}}=|m_I=+1\rangle\langle m_I=+1|\otimes\mathbbm{1}+|m_I=0\rangle\langle m_I=0|\otimes R_{-y}(\pi)$, where $\mathbbm{1}$ ($R_{-y}(\pi)$) is the evolution operator of the electron spin when the nuclear spin is in state $|m_I=+1\rangle$ ($|m_I=0\rangle$).

\begin{figure}\centering
\includegraphics[width=0.9\columnwidth]{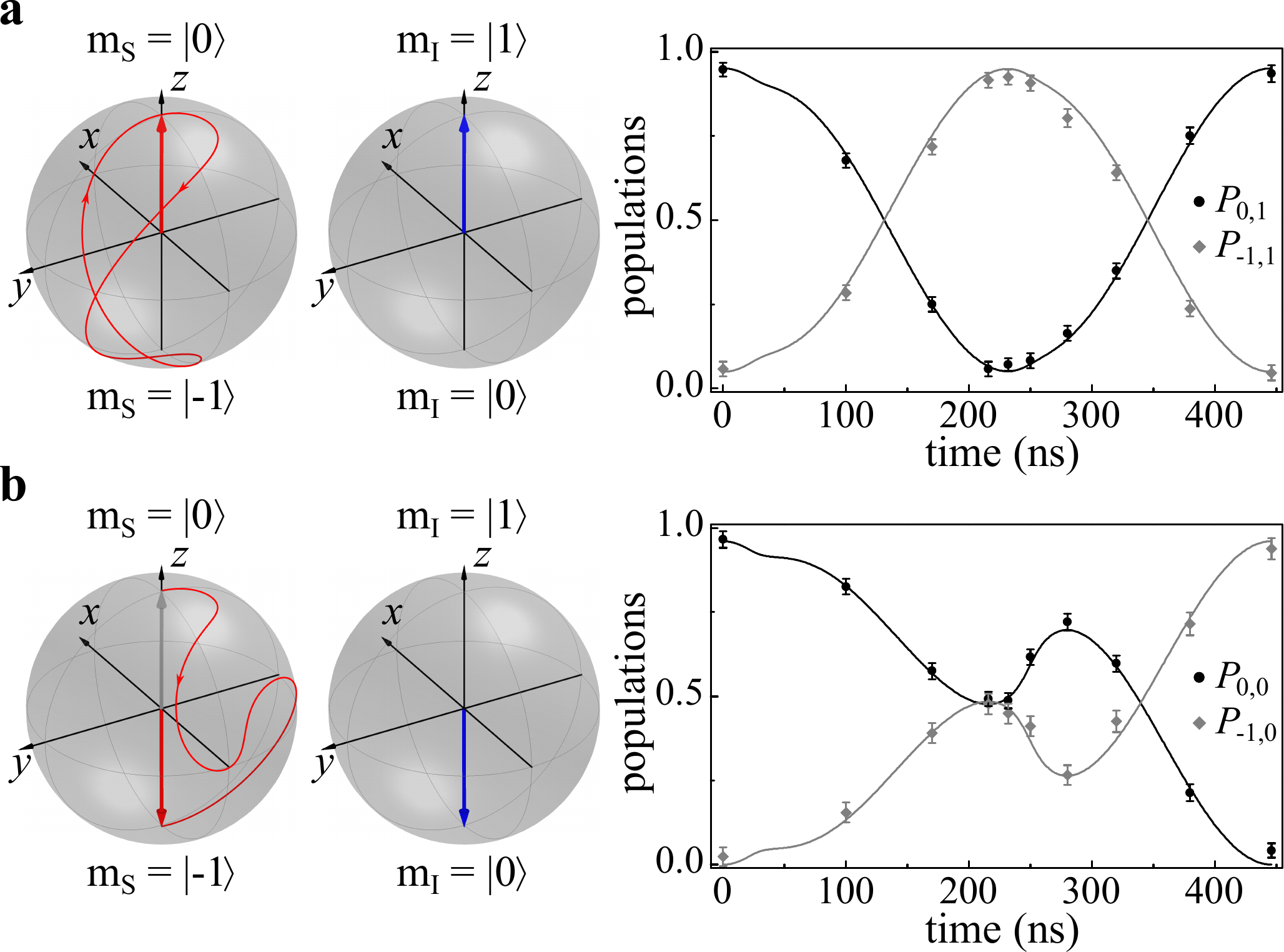}
\caption{(color online). State trajectories under the two-qubit controlled-U gate by TOC with initial states (a) $|0, 1\rangle$ and (b) $|0, 0\rangle$. The left panels show the state evolutions of the nuclear and electron spins on the Bloch spheres. When the nuclear spin qubit is in the state $|1\rangle$ ($|0\rangle$) labeled by the blue arrows, the electron spin qubit undergoes the paths labeled by red lines to the state $|0\rangle$ ($|-1\rangle$). The right panels show the experimental dynamics of state populations $P_{0,m_I}$ (black circles) and $P_{-1,m_I}$ (grey diamonds), and lines are theoretical predictions of the populations.
%    Note that due to imperfect polarization of the electron spin, there is a fraction of about 5\% population in $|-1, 1\rangle$ for the initial states.
  The error bars on the data points are the standard deviations from the mean.
 }
    \label{Fig3}
\end{figure}

Figure \ref{Fig3} shows the state evolutions under $U_c$ via TOC.
In Fig.~\ref{Fig3}a and b, the initial states are prepared into $|0, 1\rangle$ and $|0, 0\rangle$, respectively.
%An energy level diagram showing schematically the initial and final states corresponding to FIG.~\ref{Fig3}a and b is presented in Section \uppercase\expandafter{\romannumeral2} in Supplementary Material.
The left panel of Fig.~\ref{Fig3}a (b) shows the state trajectory of the electron spin qubit on the Bloch sphere, while the nuclear spin state is $|m_I=+1\rangle$ ($|m_I=0\rangle$).
It is clear that the electron spin qubit is flipped to the state $|-1\rangle$ with the nuclear spin qubit in $|m_I=0\rangle$, and the state of the electron spin qubit returns to the state $|0\rangle$ with the nuclear spin qubit in $|m_I=+1\rangle$.
In the right panels of Fig.~\ref{Fig3}a and b, experimental populations of $|0, m_I\rangle$ and $|-1, m_I\rangle$ (i.e., $P_{0, m_I}$ and $P_{-1, m_I}$)  during the $U_c$ gate are recorded.
The experimental results represented by symbols are in agreement with theoretical predictions represented as lines.
The small deviation from $1$ ($0$) of $P_{0, m_I}$ ($P_{-1, m_I}$) at $t = 0$ is due to imperfect polarization of the electron spin (about 0.95, which is measured with sequences described in Section \uppercase\expandafter{\romannumeral4} and Fig. S6 in Supplementary Material).

\begin{figure}\centering
\includegraphics[width=0.9\columnwidth]{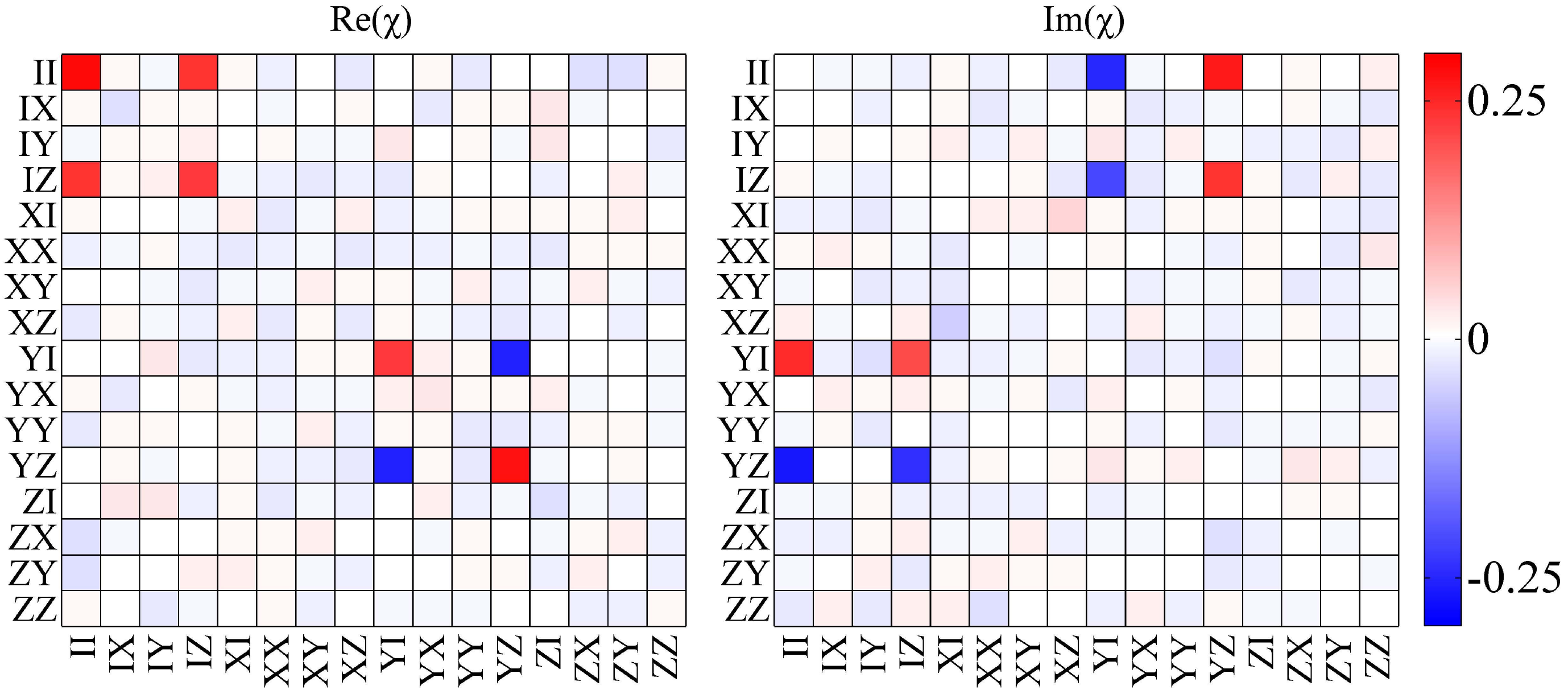}
\caption{(color online). Quantum process tomography for controlled-U gate by TOC. The left and right panels are the real and imaginary parts of the reconstructed process matrix $\chi$. The error bar of each point is about $0.01$ due to the statistics of photon counts. An average gate fidelity of 0.99(1) can be obtained from the process matrix.
 }
    \label{Fig4}
\end{figure}

We further perform quantum process tomography (see Section \uppercase\expandafter{\romannumeral3} in Supplementary Material) to characterize the $U_c$ gate. A set of 16 initial states is prepared, after which the $U_c$ is applied, and quantum state tomography is applied to reconstruct the final state corresponding to each initial state. With the information of the
$16$ final states, the process matrix $\chi$ is determined in the Pauli basis $\{\sigma_i\otimes\sigma_j\}$, where $\sigma_{i(j)}\in\{I, X, Y, Z\}$, $I$ is the identity operator, and $X=\sigma_x$, $Y=\sigma_y$, and $Z=\sigma_z$ are Pauli operators.
Figure \ref{Fig4} shows the real and imaginary parts of the experimental process matrix.
The average gate fidelity of the two-qubit gate in our experiment is $0.99(1)$, which reaches the threshold of fault-tolerant quantum computations \cite{Nature_Barends}.
The shortest possible time duration of the gate operation by TOC is advantageous to high fidelity due to the reduction of the dephasing effect.
The relatively small strength of the control field also contributes to the high fidelity, as the  noise induced from the control field is proportional to the control field \cite{NatCommun_Du,PRL_Du}.

%\section*{Discussion}
\textit{Discussion.}---Manipulation of quantum systems is of fundamental significance in quantum computing \cite{Nielsen}, quantum metrology \cite{PRL_Giovannetti},  and high-resolution spectroscopy \cite{Nature_Maze,Nature_Balasubramanian,Science_Schmidt_2005}.
It is desirable to achieve universal control with high fidelity and in a minimal time interval in the presence of decoherence.
High fidelity universal control has been reported in various quantum systems, including trapped ions \cite{Natphy_Benhelm}, superconducting circuits \cite{Nature_Barends}, NV centers in diamond \cite{Nature_Hanson,NatCommun_Du}, and spins in silicon \cite{Nature_Veldhorst,Nature_Pla}.
However, experimental demonstration of universal control, when high fidelity and minimal time are satisfied simultaneously, were not achieved in previous work.
We have realized the time-optimal universal control of the two-qubit system in diamond with high fidelity.
Our results provide an experimental validation of TOC casting a high-fidelity control operation on multi-qubit systems.
The approach developed in this work to realize accurate minimum time control of multiqubits can be applied to other important physical systems.
%including quantum dots, phosphorous doped in silicon, trapped ions and superconducting circuits.

%\begin{acknowledgments}
We are grateful to C. K. Duan and C. Y. Ju for valuable discussions.
This work was supported by the National Basic Research Program of China (Grants No.\ 2013CB921800 and No.\ 2016YFB0501603), the National Natural Science Foundation of China (Grants No.\ 11227901, No.\ 31470835, and No.\ 11275183) and the Strategic Priority Research Program (B) of the CAS (Grant No. XDB01030400). F.S. and X.R. thank the Youth Innovation Promotion Association of Chinese Academy of Sciences for support. X. W. is supported by NSF project No. CCF-1350397.

J. G. and Y.  W. contributed equally to this work.

%\begin{addendum}

%\end{addendum}

%\newpage

\onecolumngrid
\vspace{1.5cm}
\begin{center}
\textbf{\large Supplmentary Material}
\end{center}

\setcounter{figure}{0}
\setcounter{equation}{0}
\makeatletter
\renewcommand{\thefigure}{S\@arabic\c@figure}
\renewcommand{\theequation}{EqS\@arabic\c@equation}
\renewcommand{\bibnumfmt}[1]{[RefS#1]}
\renewcommand{\citenumfont}[1]{RefS#1}

\def\U{\mathrm{U}}
\def\A{\mathcal{A}}
\def\B{\mathcal{B}}
\def\C{\mathcal{C}}
\def\D{\mathcal{D}}
\def\E{\mathcal{E}}
\def\F{\mathcal{F}}
\def\G{\mathcal{G}}
\def\H{\mathcal{H}}
\def\I{\mathcal{I}}
\def\L{\mathcal{L}}
\def\M{\mathcal{M}}
\def\N{\mathcal{N}}
\def\O{\mathcal{O}}
\def\P{\mathcal{P}}
\def\Q{\mathcal{Q}}
\def\T{\mathcal{T}}
\def\R{\mathbb{R}}
\def\J{\mathcal{J}}
\def\S{\mathcal{S}}
\def\K{\mathcal{K}}
\def\Tr{\operatorname{Tr}}

\section{\uppercase\expandafter{\romannumeral1}. Hamiltonian of the NV system}
%\subsection{Hamiltonian of the NV system}
\textrm{\\}
The Hamiltonian of the NV center can be  written as
\begin{equation}
H_{NV}=2\pi(DS_{z,3}^2+\omega_SS_{z,3}+PI_{z,3}^2-\omega_II_{z,3})+H_{hf},
\end{equation}
where $\omega_S=-\gamma_eB_0/2\pi(\omega_I=\gamma_NB_0/2\pi)$ is the Zeeman splitting of the electron ($^{14}$N nuclear) spin, $\gamma_e (\gamma_N)$ is the electronic ($^{14}$N nuclear) gyromagnetic ratio, $S_{z,3}$ and $I_{z,3}$ are the electron and nitrogen nuclear spin operators of spin-1 systems. The zero field splitting $D = 2870$ MHz and the nuclear quadrupolar splitting $P = -4.95$ MHz. The hyperfine interaction between the NV electron and the $^{14}$N nuclear spin is
\begin{equation}
H_{hf}=2\pi[A_{\perp}(S_{x,3}I_{x,3}+S_{y,3}I_{y,3})+AS_{z,3}I_{z,3}],
\end{equation}
where $A = -2.16$ MHz. Because of the strong zero field splitting and Zeeman splitting terms of the electron spin, the effect of the interaction term $S_{x,3}I_{x,3}+S_{y,3}I_{y,3}$ can be neglected. In the secular approximation, the Hamiltonian is
\begin{equation}
H_{NV}=2\pi(DS_{z,3}^2+\omega_SS_{z,3}+AS_{z,3}I_{z,3}+PI_{z,3}^2-\omega_II_{z,3}),
\end{equation}

\begin{figure}[!h]\centering
\includegraphics[width=0.5\columnwidth]{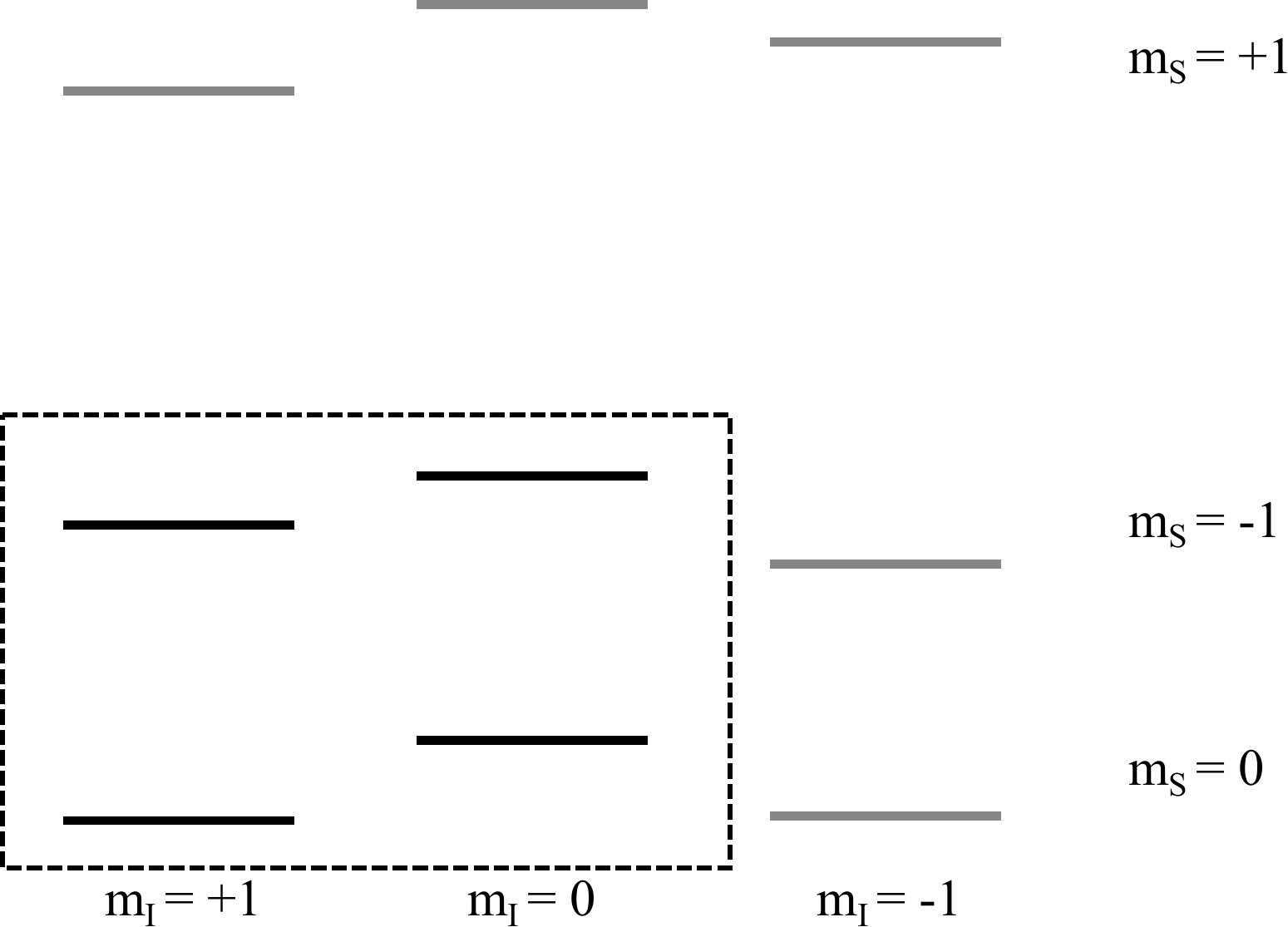}
\caption{\label{NVEnergyLevel} \textbf{Spin energy level diagram of the NV center.} The experiments are implemented on the two-qubit system composed with the energy levels boxed with dashed lines (i.e. $|m_S=0, m_I=1\rangle$, $|m_S=0, m_I=0\rangle$, $|m_S=-1, m_I=1\rangle$, and $|m_S=-1, m_I=0\rangle$).
}
\end{figure}

The spin energy levels of the NV center are shown in Fig.~\ref{NVEnergyLevel}.
The electron (nuclear) spin states $|m_S = 0\rangle$ and $|m_S = -1\rangle$ ($|m_I = 1\rangle$ and $|m_I = 0\rangle$) are encoded as the electron (nuclear) spin qubit.
The Hamiltonian can be simplified to that of a two-qubit system.

In the single-qubit case, the experiments are implemented on the electron spin qubit while the nuclear spin is kept in state $|m_I=1\rangle$. When microwave (MW) pulses with the frequency of $f_{MW}$ are applied, the total Hamiltonian of the electron spin qubit is
\begin{equation}
H_1=2\pi(-(D-\omega_S-A)S_z+2\nu_1\cos(2{\pi}f_{MW}t-\phi)S_x),
\end{equation}
where $\phi$ is the phase of the MW pulse, $\nu_1$ is the amplitude of the MW pulse, $S_x$, $S_y$, and $S_z$ are electron spin operators of a spin-1/2 system.
The Hamiltonian can be transformed into the rotating frame as
\begin{equation}
H_{rot1}=U_{trans1}H_1U_{trans1}^{\dagger}-iU_{trans1}\frac{dU_{trans1}^{\dagger}}{dt},
\end{equation}
with
\begin{equation}
U_{trans1}=e^{-i2{\pi}f_{MW}tS_z}.
\end{equation}
With rotating-wave approximation, the Hamiltonian in the rotating frame can be simplified as
\begin{equation}
H_{rot1}=2{\pi}({\delta_0}S_z+\nu_1(\cos\phi S_x+\sin\phi S_y)),
\end{equation}
where $\delta_0 = -(D - \omega_S - A - f_{MW})$.

The two-qubit experiments are implemented on the two-qubit system comprised with the electron and nuclear spin qubits. The two-qubit states can be manipulated with MW and radio-frequency (RF) pulses.
The frequency of the RF pulse is denoted by $f_{RF}$. When only MW pulses are applied, the total Hamiltonian is
\begin{equation}
H_{2}=2\pi(-(D-\omega_S-A/2)S_z+(P-\omega_I-A/2)I_z+AI_zS_z+2\nu_{1}\cos(2{\pi}f_{MW}t-\phi)S_x),
\end{equation}
where $I_x$, $I_y$, $I_z$ are nuclear spin operators of a spin-1/2 system.
The Hamiltonian can be transformed into the rotating frame as
\begin{equation}
H_{rot2}=U_{trans2}H_{2}U_{trans2}^{\dagger}-iU_{trans2}\frac{dU_{trans2}^{\dagger}}{dt},
\end{equation}
with
\begin{equation}
U_{trans2}=e^{-i2{\pi}f_{MW}tS_z}e^{-i2{\pi}f_{RF}tI_z}.
\end{equation}
With rotating-wave approximation, the Hamiltonian in the rotating frame can be simplified as
\begin{equation}\label{eqn:ham-nv}
H_{rot2}=2\pi[AS_zI_z+\nu_{1}(\cos{\phi}S_x+\sin{\phi}S_y)],
\end{equation}
with $f_{MW} = D-\omega_S-A/2$, $f_{RF} = -P+\omega_I+A/2$.

%\section{Quantum brachistochrone equation for realistic system}
\section{\uppercase\expandafter{\romannumeral2}. Time-optimal control with quantum brachistochrone equation}
%\subsection{Quantum brachistochrone equation for realistic system}
\textrm{\\}
The goal of quantum time-optimal control (TOC) is to complete the quantum control task(e.g., to generate a unitary gate or to prepare an entangled state) in the shortest time. In general, there are two constraints for the quantum TOC problem: (i) due to the finite energy bandwidth of the quantum system, its evolution under the Schr\"odinger equation cannot be arbitrarily fast; (ii) the Hamiltonian of a realistic quantum device usually takes a given form, and the way we can vary the Hamiltonian must satisfy this form, implying we cannot generate arbitrary quantum trajectory. Under these two constraints,  it has been found that the quantum TOC problem can be solved by solving the so-called quantum brachistochrone equation (QBE)~\cite{PRL_Carlini_S,PRA_Carlini_S}.

Specifically, let $H(t)$ be the Hamiltonian of the quantum device that can be varied over time. In the most general case, $H(t)=H_0+H_c(t)$, where $H_0$ is known as the drift term which cannot be varied over time (e.g., $H_0=2\pi AS_zI_z$ in (\ref{eqn:ham-nv})), and $H_c$ is the control Hamiltonian in a fixed form (e.g., $H_c(t)=2\pi \nu_{1}(\cos{\phi(t)}S_x+\sin{\phi(t)}S_y)$ in (\ref{eqn:ham-nv})). Then the above two constraints can be expressed as the following: \\
(i) $f_0(H)\equiv||H_c(t)||^2-E^2=\Tr(H_c^2(t))-E^2=0$\\
(ii) $ f_k(H)\equiv\Tr(H_c(t)B_k)=0, \text{ where } k=1,\cdots,m$\\
where $\{B_k\}$ is a basis of the matrix subspace $\Q$ satisfying $\Tr(H_c(t)\Q)=0$. Notice that, in principle, the constraint for finite energy bandwidth should be expressed as the inequality (i') $||H_c(t)||^2\le E^2$, but we find that (i') leads to (i) in many cases, including the problems discussed in this work.

For the gate generation problem, the objective is to find the appropriate control Hamiltonian $H_c(t)$ such that the evolution $U(t)$ under the Schr\"{o}dinger equation $\dot{U}=-iH(t)U$ satisfies $U(0)=I$ and $U(T)=U_F$, where $U_F$ is the target unitary and $T$ is the total evolution time. This control solution $(H(t),U(t))$ is not unique, and numerically we can use optimization method to find many such solutions. However, if the control objective also requires the total time $T$ to be minimized, then the time-optimal control can be mathematically characterized: defining the Lagrangian $\L$ under the constraints (i) and (ii), the time-optimal solution $(H(t),U(t))$ must satisfy the Euler-Lagrange equation for $\L$:
\begin{equation}
\label{QBE}
\dot{F}=-i[H,F],
\end{equation}
where $L_c\equiv\sum_j\lambda_jf_j(H)$, $j=0,1\cdots m$, $\lambda_j$ are Lagrange multipliers and $F=\frac{\partial L_c}{\partial H}=\frac{\partial L_c}{\partial H_c}$ since $H_0$ is constant.

Since the QBE is a first-order ordinary differential equation (ODE), in order to solve the time-optimal gate generation problem for $U_F$, it is sufficient to find the initial value $(H(0),\lambda_j(0))$ satisfying $U(T)=U_F$. This is a boundary value nonlinear problem. Except a few special cases where analytic solutions exist, one has to resort numerical method to solve it. Unfortunately, the standard method (e.g. the shooting method) of numerically solving a boundary value nonlinear equation soon becomes inefficient as the dimension of the problem grows. This forces us to think of new method to solve the QBE.

Fortunately, for the special case where $H_0=0$, the QBE has an intuitive geometric interpretation, i.e., it can be considered as the limit of a family of 1-parameter  geodesics under the so-called $q$-metric~\cite{PRL_Wang_S}. Such brachistochrone-geodesic connection provides a very efficient way of solving the QBE even for the system  with a large dimension. Analogously, for the more general case where $H_0\neq 0$, we can develop a similar method to solve the QBE.

Specifically, let $\P$ be the matrix subspace where $H_c(t)$ can choose values in, so we have $\Tr(\P\Q)=0$. Notice that the key point to solve the first-order QBE (\ref{QBE}) is to find a good guess of the initial value $(H_c(0),\lambda_j(0))$. This can be achieved by the following approach:

First, assume $H_c(t)$ is only chosen from the control space $\P$. The time-optimal problem is equivalent
to solving the following two-objective optimization problem: maximizing the fidelity of evolution
operator with the target operator, $\mathrm{Fi}(U(T),U_F)$, and minimizing the evolution time, $T=\int_0^T dt$.
Meanwhile, the control Hamiltonian is subjected to the constraint $||H_c(t)|| = E$. In order to solve
it, we can use weighted summation to convert the two-objective problem into a single-objective
minimization problem ($\kappa>0$), with the objective function:
\begin{align}\label{objfunction}
&J=-\mathrm{Fi}(U(T),U_F)+\kappa \int_0^T dt
\end{align}
We expect that this combined optimization problem will give us a reasonably good approximation of
time-optimal solution $H_c(t)$. Thus, we can get a good guess for $H_c(0)$, but this is not sufficient to solve the QBE, as the initial values for the Lagrange multipliers $\lambda_j(0)$ are still unknown.

To overcome this problem, we introduce a family of 1-parameter brachistochrone equations under the $q$-metric, which is similar to the family of the geodesic equations in Ref.~\onlinecite{PRL_Wang_S}. In this frame, the time-optimal curves are reformulated in the following way: allowing the
control Hamiltonian $H_c(t)$ to take components from both $\P$ and $\Q$ but with a penalty when taking
components from $\Q$. Specifically, we define the $q$-norm, characterizing the penalty:  $||H_c||_q^2=||\mathcal{P}(H_c)||^2+q||\mathcal{Q}(H_c)||^2$, $q\ge 1$, where $\mathcal{P}(H_c)$ ($\mathcal{Q}(H_c)$) indicate the projection of $H_c$ in subspace $\mathcal{P}$ ($\mathcal{Q}$).
Then we study the time-optimal solution under the following constraint: $||H_c||_q=E$. As $q\to +\infty$, the component of $H_c(t)$ on the subspace $\mathcal{Q}$ decreases to zero, we will recover the brachistochrone equation for the original problem. Thus, we can solve the weighted-sum optimization problem shown in (\ref{objfunction}), with the constraint $||H_c(t)||_q = E$. The optimal solution provides a good initial guess of $(H_c(0),q\lambda_j(0))$, which can be used to find the solution of the $q$-metric brachistochrone equation for $q\gg 1$, which then provides a good guess to solve the original QBE. The detailed procedure is similar to that in the Ref.~\onlinecite{PRL_Wang_S}.
%The detailed procedure can be found in the Ref.~\onlinecite{PRL_Wang}.

\subsection{A. Single-qubit case}
%We first consider the case that the control Hamiltonian $H_c(t)$ is allowed to take components from the entire Hilbert space (i.e., $H_c(t){\in}\mathrm{span}\{S_x,S_y,S_z\}$).
%We give the solutions of the target operator $U_F=e^{-i{\theta}S_z}$ (trivial case) and $U_F=e^{-i{\theta}S_x}$ (non-trivial case).
%Experimental demonstration is shown in the Supplementary Table \ref{siglequibitresults}.
The entire Hamiltonian is $H(t)=H_0+H_c(t)$, with $H_0=2\pi{\delta}S_z$ and $H_c(t)=2\pi\nu_{1}[\cos\phi(t)S_x+\sin\phi(t)S_y]$. The target operator is $U_F=R (\hat {\mathbf n} ,\theta) \equiv \exp(-i\theta\hat {\mathbf n}\cdot\textbf{S})$, with $\hat {\mathbf n}=\hat {\mathbf x}\sin\gamma\cos\varphi+\hat {\mathbf y}\sin\gamma\sin\varphi + \hat {\mathbf z}\cos\gamma$ being a unit vector.
From QBE, we have $\phi(t)=2\pi\eta t+\phi_0$. The parameter $\eta$ characterizing the control Hamiltonian and the minimum evolution time $T$ can be obtained by solving the Schr\"{o}dinger equation with boundary conditions.
The solution to the Schr\"{o}dinger equation reduces the boundary conditions to the following equations.

\begin{equation}
\left\{
\begin{aligned}
&\cos(\pi\eta T)\cos(\pi\sqrt{\nu_{1}^2+(\delta-\eta)^2}T)\\
&~~~~~~~-\sin(\pi\eta T)\sin(\pi\sqrt{\nu_{1}^2+(\delta-\eta)^2}T)\frac{\delta-\eta}{\sqrt{\nu_{1}^2+(\delta-\eta)^2}}=\pm\cos\frac{\theta}{2}\\
&\sin(\pi\eta T)\cos(\pi\sqrt{\nu_{1}^2+(\delta-\eta)^2}T)\\
&~~~~~~~+\cos(\pi\eta T)\sin(\pi\sqrt{\nu_{1}^2+(\delta-\eta)^2}T)\frac{\delta-\eta}{\sqrt{\nu_{1}^2+(\delta-\eta)^2}}=\pm\cos\gamma\sin\frac{\theta}{2}\\
&\frac{\nu_{1}}{\sqrt{\nu_{1}^2+(\delta-\eta)^2}}\cos(\pi\eta T+\phi_0)\sin(\pi\sqrt{\nu_{1}^2+(\delta-\eta)^2}T)=\pm\sin\gamma\cos\varphi\sin\frac{\theta}{2}\\
&\frac{\nu_{1}}{\sqrt{\nu_{1}^2+(\delta-\eta)^2}}\sin(\pi\eta T+\phi_0)\sin(\pi\sqrt{\nu_{1}^2+(\delta-\eta)^2}T)=\pm\sin\gamma\sin\varphi\sin\frac{\theta}{2}
\end{aligned}
\right.
\end{equation}

%We take realizing $U_F = \exp (-i{\theta}S_z)$ as a nontrivial case, because the control Hamiltonian is restricted ($H_c(t){\in}\mathrm{span}\{S_x,S_y\}$).
%The analytic solution when $U_F = \exp (-i{\theta}S_z), \theta\in(0,2\pi)$ is shown in FIG.~\ref{analytic solution}.
As an example, we give analytic solution to $U_F = R (\hat {\mathbf z} ,\theta), \theta\in(0,2\pi)$. Without loss of generality, $\delta\geq 0$ and $\nu_1>0$ is supposed.

\begin{figure}[!b]\centering
\includegraphics[width=0.7\columnwidth]{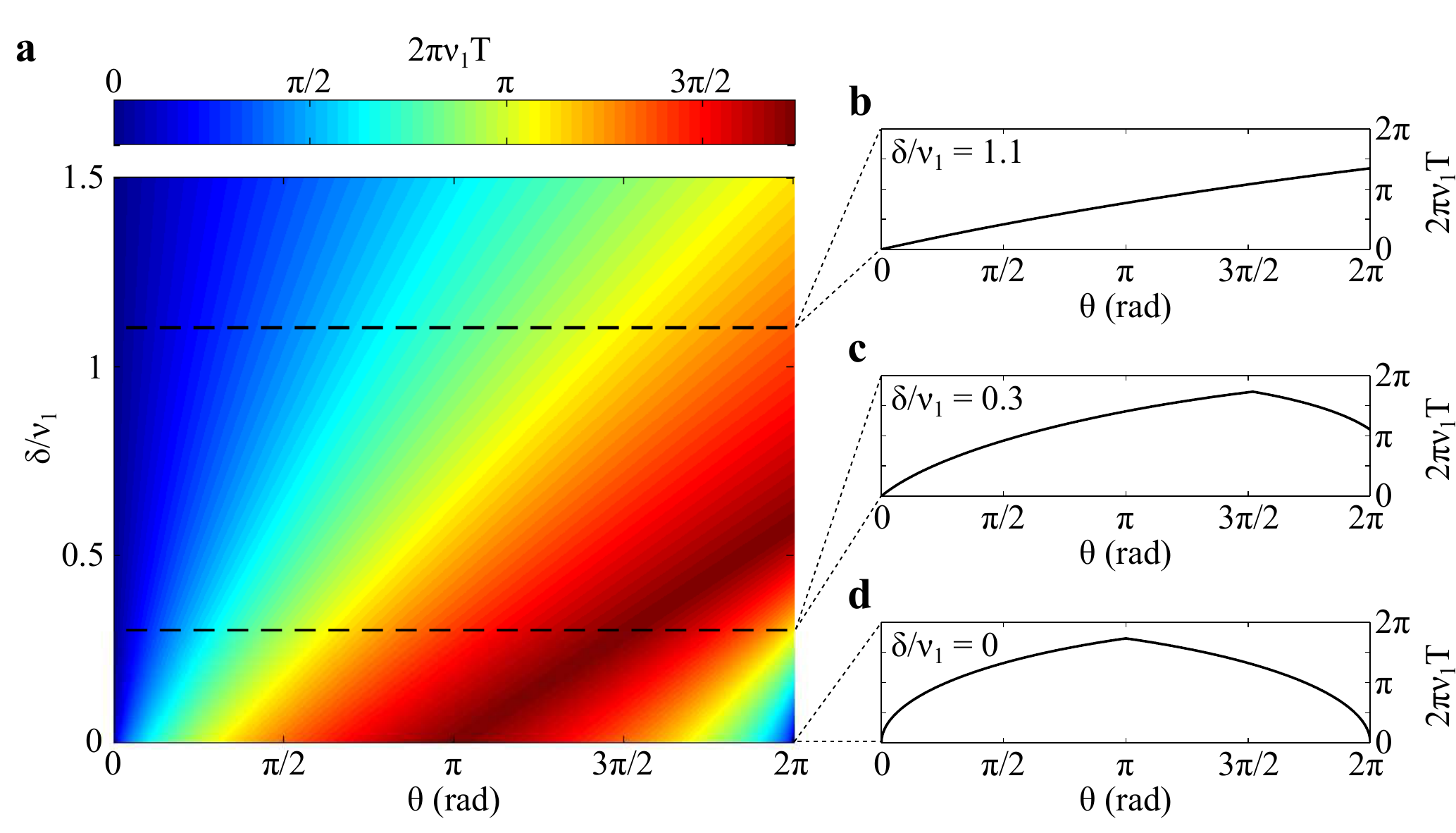}
\caption{\label{analytic solution} \textbf{Analytic solution of single-qubit TOC for target evolution operator $R (\hat {\mathbf z} ,\theta)$.} (a) The minimum evolution time analytically calculated from QBE with $H_c\in\mathrm{span}\{S_x, S_y\}$ and target evolution operator $R (\hat {\mathbf z} ,\theta)$ with $\theta \in [ 0, 2\pi )$. (b - d) show the minimum evolution time as a function of rotation angle for different ratios between $\delta$ and $\nu_1$.}
\end{figure}

When $\theta <  \pi (1 + \sqrt{3} \delta / \nu_1)$
%$\delta\ge\frac{1}{\sqrt{3}}\nu_{1}$ or $\delta<\frac{1}{\sqrt{3}}\nu_{1}, \theta\le(1+\sqrt{3}\frac{\delta}{\nu_{1}})\pi$,
\begin{equation}
\label{Eq_T_1}
T=\frac{\delta(\frac{\theta}{2\pi}-1)+\sqrt{\nu_{1}^2+\delta^2-\nu_{1}^2(\frac{\theta}{2\pi}-1)^2}}{\nu_{1}^2+\delta^2}
\end{equation}
\begin{equation}
\eta=-\frac{2\pi-\theta}{2\pi T}.
\end{equation}

When $\theta\geq\pi(1+\sqrt{3}\frac{\delta}{\nu_{1}})$
%$\delta<\frac{1}{\sqrt{3}}\nu_{1}, \theta>(1+\sqrt{3}\frac{\delta}{\nu_{1}})\pi$,
\begin{equation}
\label{Eq_T_2}
T=\frac{\delta\frac{\theta}{2\pi}+\sqrt{\nu_{1}^2+\delta^2-\nu_{1}^2(\frac{\theta}{2\pi})^2}}{\nu_{1}^2+\delta^2}
\end{equation}
\begin{equation}
\eta=-\frac{\theta}{2\pi T}.
\end{equation}

The minimum time $T$ as a function of $\theta$ and $\delta/\nu_1$ (equation \ref{Eq_T_1} and \ref{Eq_T_2}) is shown in Fig. \ref{analytic solution}.
When $\delta=0$, our result reduces to that in Ref.~\onlinecite{PRA_Boozer_S}.
In this case, we compare the time duration with TOC and that with Euler rotation.
The experimental results is shown in the main text.
%We implement experiments to compare the time cost between the derived TOC and a non-optimized evolution path.
%The realization of a target $R (\hat {\mathbf z} ,\theta), \theta\in(0,\pi)$ when $\delta=0$ is taken as an example.
%A typical realization is according to the Euler rotation:

\begin{figure}[!h]\centering
\includegraphics[width=0.5\columnwidth]{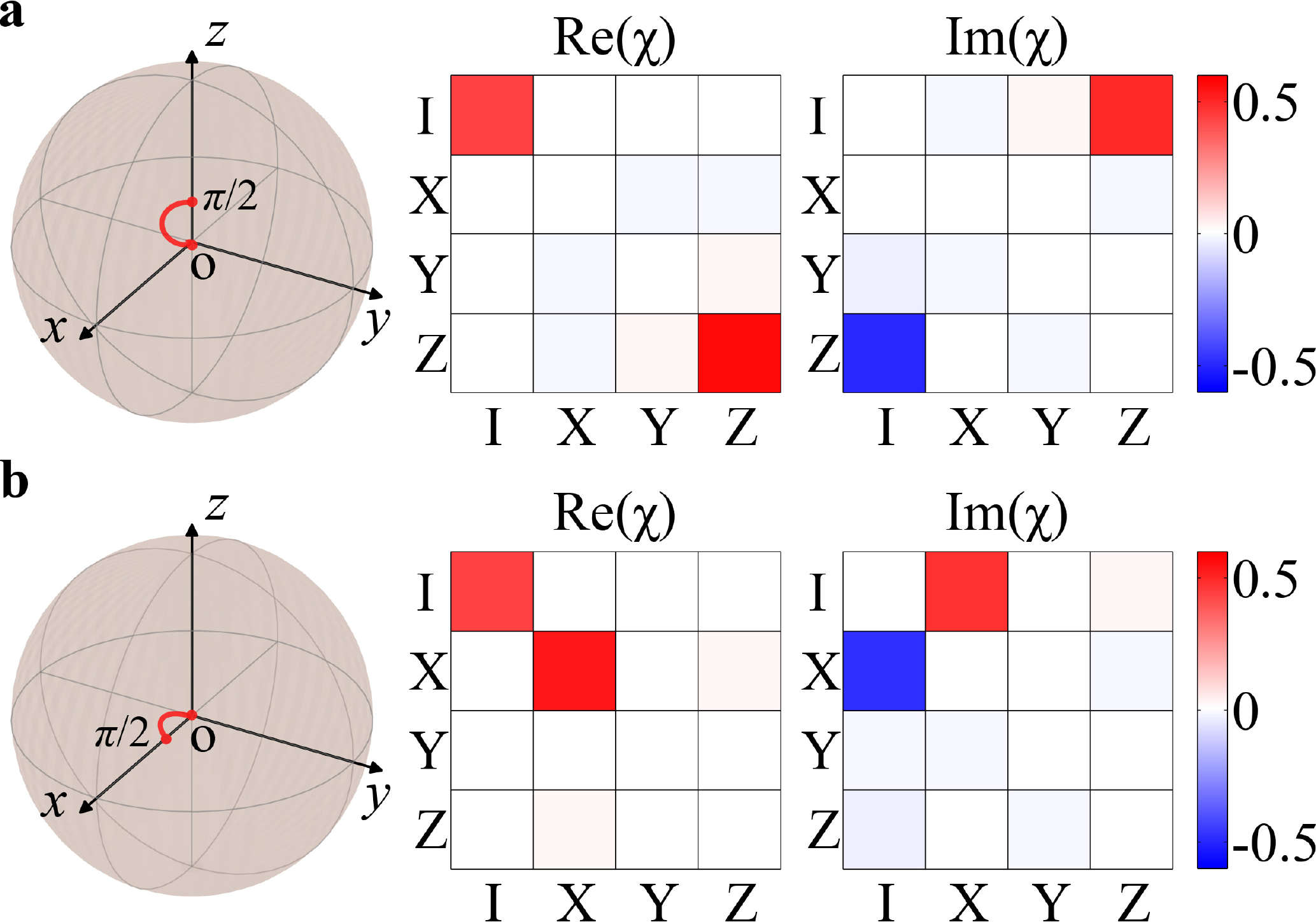}
\caption{\label{single_qubit_tomo}\textbf{ Experimental implementation of single-qubit TOC for target evolution operators} (a) $R (\hat {\mathbf z} ,\pi/2)$ and (b) $R (\hat {\mathbf x} ,\pi/2)$. The left panels show the time-optimal evolution paths. The spheres in the left column stand for the rotation group\cite{PRL_Meier_S}. The center of the sphere O stands for the identity gate operation. The direction of the vector joining the center O to any other point represents the axis of rotation, whereas the length of the vector represents the angle of rotation. The right panels are experimental operations characterized by quantum process tomography. The results show average gate fidelities of (a) 1.00(1) and (b) 0.99(1), respectively.
 }
\end{figure}

Then we experimentally demonstrate the case of $\delta\not=0$.
In this experiment we set $\delta=1.5~$MHz and $\nu_1=5~$MHz.
To visualize the evolution path of the TOC, the evolution operator $R (\hat {\mathbf n} ,\theta)$ is mapped to the point $(\theta\sin\gamma\cos\varphi, \theta\sin\gamma\sin\varphi, \theta\cos\gamma)$ in a three-dimensional frame of $\theta\in[0, 2\pi]$.
All single-qubit evolution operators can be mapped to points within the sphere  with radius of $2\pi$ \cite{PRL_Meier_S}.
The time-optimal evolution paths of the TOC for target evolution operators $R (\hat {\mathbf z} ,\pi /2)$  and $R (\hat {\mathbf x} ,\pi /2 )$ are represented in the left panels of Fig.~\ref{single_qubit_tomo}a and b, respectively.
To characterize the performance of the TOC, quantum process tomography \cite{Nielsen_S} is utilized (see Section \uppercase\expandafter{\romannumeral3}).
%A physical process can be represented by a completely positive map $\varepsilon$ on an arbitrary input state $\rho$: $\varepsilon(\rho)=\sum_{mn}\chi_{mn}A_m\rho A_n^\dag$, with operators ${A_n}$ forming a basis and $\chi$ being the process matrix.
%We determine the process matrix $\chi$ in the Pauli basis $A_n=\{I, X, Y, Z\}$, where $I$ is the identity operator, and $X=\sigma_x$, $Y=\sigma_y$, and $Z=\sigma_z$ are Pauli operators.
The reconstructed process matrices for $R (\hat {\mathbf z} ,\pi / 2)$ and $R (\hat {\mathbf x} ,\pi / 2)$ are shown in the right panels of Fig.~\ref{single_qubit_tomo}a and b, respectively.
The corresponding average gate fidelities \cite{PLA303_249_S} are measured to be $1.00(1)$ and $0.99(1)$.

\begin{figure}[!h]\centering
\includegraphics[width=0.7\columnwidth]{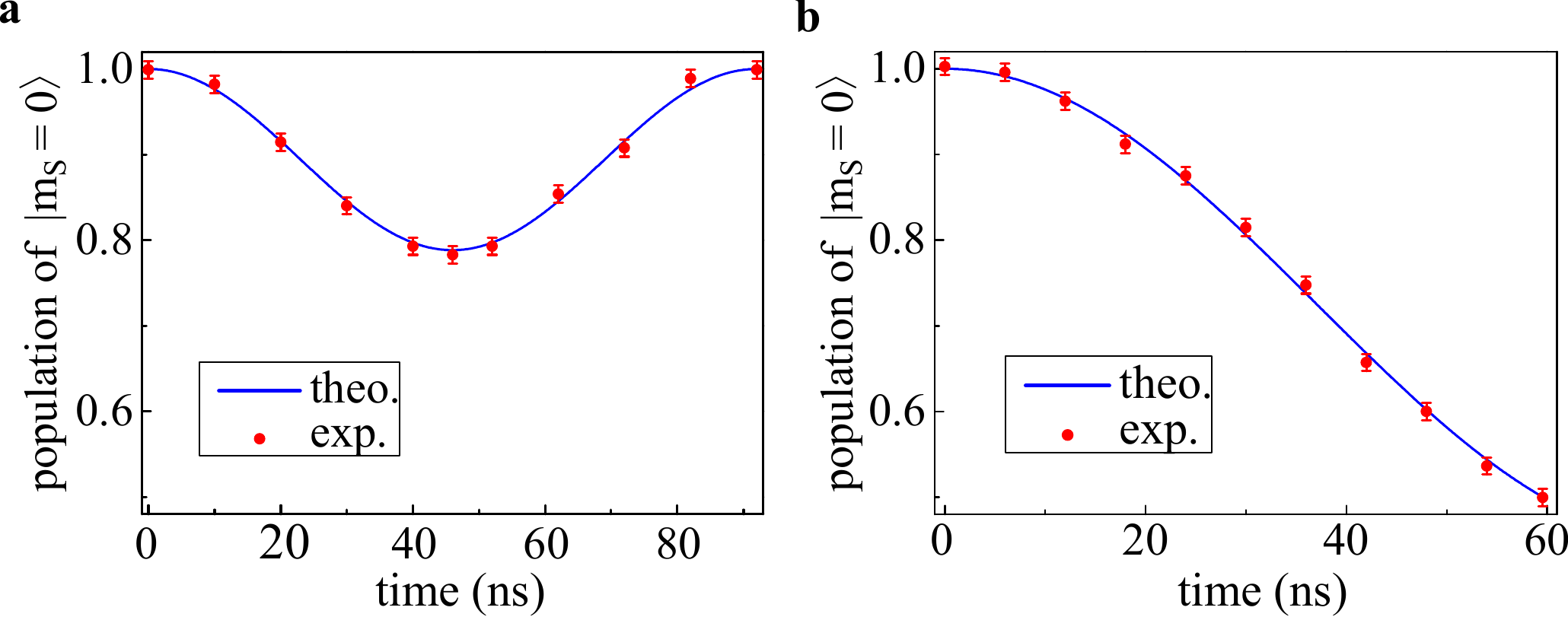}
\caption{\label{single-qubit state evolution} \textbf{State evolution under (a) $R (\hat {\mathbf z} ,\pi /2)$  and (b) $R (\hat {\mathbf x} ,\pi /2 )$ by single-qubit TOC with initial state $|m_s=0\rangle$.} The control Hamiltonian $H_c\in\mathrm{span}\{S_x, S_y\}$, with other parameters set to be $\delta=1.5$~MHz, $\nu_1=5$~MHz. The symbols are experimentally recorded state population of $|m_S=0\rangle$, and solid lines are theoretical predictions. The error bars on the data points are the standard deviations from the mean.}
\end{figure}

We further exhibit the state evolutions during the time-optimal $R (\hat {\mathbf z} ,\pi /2 )$ and $R (\hat {\mathbf x} ,\pi /2 )$.
%We now focus on the experimental implementation of single-qubit time-optimal gate operators when $\delta>0$.
The initial state is prepared to $|m_S=0\rangle$.
State populations of $|m_S=0\rangle$ during the evolutions are recorded.
As shown in Fig. \ref{single-qubit state evolution}, experimental results are in great agreement with theoretical predictions.
%We also perform quantum process tomography to characterize the realized operators.
%The experimental results are provided in the main text.

%We take target evolution operators $R (\hat {\mathbf z} ,\pi /2)$  and $R (\hat {\mathbf x} ,\pi /2 )$ as examples to show the state evolution under TOC in FIG.~\ref{single-qubit state evolution}.
%The initial electron spin-qubit states are prepared into $|0\rangle$, respectively.
%It is clear that electron spin-qubit return to state $|0\rangle$ under operator $R (\hat {\mathbf z})$, and result in the state in XY-plane under operator $R (\hat {\mathbf x} ,\pi /2 )$.
%The experimental population of $|0\rangle$ represented by symbols are in agreement with theoretical predictions represented as lines.
%The result of quantum process tomography which characterize the performance of the TOC is shown in the main text.

In addition to the experiments mentioned above, more single-qubit gate operators with TOC are experimentally demonstrated.
Our method also applies to TOC with other constraints on $H_c$ (e.g. $H_c(t){\in}\mathrm{span}\{S_x,S_y,S_z\}$).
We also implement single-qubit TOC with constraint $H_c(t){\in}\mathrm{span}\{S_x,S_y,S_z\}$.
All the experimental results are summarized in TABLE~\ref{siglequibitresults}.
Our results show the universality of our approach to perform time-optimal universal control for single qubit.

%We experimental demonstrate more different rotation angles about $\hat z$.
%Except for the various of rotation axe and angle we also give the result of different drift Hamiltonian.
%Our method can also provide the time-optimal way to realize rotations around different axes, such as $\hat x$ and $\hat y$.
%Experiments are carried out to demonstrated these operations, and the fidelities are determined by quantum process tomography.
%The resulted fidelities with different axes are also included in the TABLE~\ref{siglequibitresults}.
%Operations with various rotation angles with different axes are presented.
%Furthermore, we have also implemented time-optimal universal control of single-qubit with other $\delta$ values and constraints on $H_c$ ($H_c(t){\in}\mathrm{span}\{S_x,S_y,S_z\}$).
%These results are shown in the TABLE~\ref{siglequibitresults}.
%Our results show the universality of our approach to perform time-optimal universal control for single qubit.

\begin{table}[!h]\centering  % ±í¾ÓÖÐ
\caption{\textbf{Summarization of the results of single-qubit TOC.} Average gate fidelity ($F_{\textrm{a}}$) and evolution time ($T \mathrm{(ns)}$) is shown for different cases with experimental parameter $\nu_{1}=5$~MHz.}
\textrm{\\}
\renewcommand{\multirowsetup}{\centering}
\begin{tabular}{c|c|ccc}  % {lccc} ±íʾ¸÷ÁÐÔªËضÔÆ뷽ʽ£¬left-l,right-r,center-c
\hline\hline
$\delta/\nu_1$ & Control space & Target operator & $F_{\textrm{a}}$ & $T \mathrm{(ns)}$ \\ \hline  % \hline ÔÚ´ËÐÐÏÂÃæ»­Ò»ºáÏß
\multirow{3}{*}{$\delta/\nu_1=0$} & \multirow{3}{*}{$H_c\in\mathrm{span}\{S_x,S_y\}$} & $R (\hat {\mathbf z} ,\pi / 2)$ & 0.99(1) & 132.3 \\          % \\ ±íʾÖØпªÊ¼Ò»ÐÐ
 & & $R (\hat {\mathbf z} ,5\pi/4)$ & 0.98(1) & 156.1\\
 & & $R (\hat {\mathbf z} ,7\pi/4)$ & 0.98(1) & 96.8\\  \hline

 \multirow{10}{*}{$\delta/\nu_1=0.3$} & \multirow{5}{*}{$H_c\in\mathrm{span}\{S_x,S_y,S_z\}$} & $R (\hat {\mathbf z} ,\pi/2)$ & 0.99(1) & 38.5 \\
 & & $R (\hat {\mathbf z} ,5\pi/4)$ & 0.98(1) & 96.2\\
 & & $R (\hat {\mathbf z} ,7\pi/4)$ & 0.99(1) & 35.7\\
 & & $R (\hat {\mathbf x} ,\pi/4)$ & 0.98(1) & 26.1\\
 & & $R (\hat {\mathbf x} ,\pi/2)$ & 0.99(1) & 51.9\\ \cline{2-5}
 %& & $R_x(\pi)$ & 0.99(1) & 100\\
 &\multirow{5}{*}{$H_c\in\mathrm{span}\{S_x,S_y\}$} & $R (\hat {\mathbf z} ,\pi/2)$ & 1.00(1) & 92.0\\
 & & $R (\hat {\mathbf z} ,5\pi/4)$ & 0.98(1) & 158.1\\
 & & $R (\hat {\mathbf z} ,7\pi/4)$ & 0.99(1) & 152.7\\
 & & $R (\hat {\mathbf x} ,\pi/4)$ & 0.98(1) & 63.2\\
 & & $R (\hat {\mathbf x} ,\pi/2)$ & 0.99(1) & 59.6\\ \hline
 %& & $R_x(\pi)$ & 0.99(1) & 100\\

 \multirow{10}{*}{$\delta/\nu_1=1.1$} & \multirow{5}{*}{$H_c\in\mathrm{span}\{S_x,S_y,S_z\}$} & $R (\hat {\mathbf z} ,\pi/2)$ & 0.99(1) & 23.8 \\
 & & $R (\hat {\mathbf z} ,5\pi/4)$ & 0.99(1) & 59.5\\
 & & $R (\hat {\mathbf z} ,7\pi/4)$ & 0.99(1) & 83.3\\
 & & $R (\hat {\mathbf x} ,\pi/4)$ & 0.99(1) & 95.4\\
 & & $R (\hat {\mathbf x} ,\pi/2)$ & 0.98(1) & 96.0\\ \cline{2-5}
 %& & $R_x(\pi)$ & 0.99(1) & 100\\
 &\multirow{5}{*}{$H_c\in\mathrm{span}\{S_x,S_y\}$} & $R (\hat {\mathbf z} ,\pi/2)$ & 0.99(1) & 41.5\\
 & & $R (\hat {\mathbf z} ,5\pi/4)$ & 0.98(1) & 92.9\\
 & & $R (\hat {\mathbf z} ,7\pi/4)$ & 0.99(1) & 121.6\\
 & & $R (\hat {\mathbf x} ,\pi/4)$ & 0.98(1) & 122.9\\
 & & $R (\hat {\mathbf x} ,\pi/2)$ & 0.99(1) & 111.7\\ \hline \hline
 %& & $R_x(\pi)$ & 0.99(1) & 100\\
\end{tabular}
%\caption{Summarization of the RB results for naive, five-piece SUPCODE, BB1, and BB1inC pulses. }
\label{siglequibitresults}
\end{table}

\subsection{B. Two-qubit case}
We exhibit the approach to obtain the time-optimal control of two-qubit system in NV center.
The drift Hamiltonian is the hyperfine coupling between electron and nuclear spin qubits. The system is steered by control pulse on the electron spin-qubit with finite strength. Thus, $H_0=2\pi{A}S_zI_z$ and $H_c(t)=2\pi\nu_{1}[\cos\phi(t)S_x+\sin\phi(t)S_y]$.
The hyperfine coupling strength is $A=\mathrm{-2.16}$MHz.
The strength of the control field $\nu_1$ is set to 2.5 MHz. The constraints on the control Hamiltonian can be described by $f_0(H)=0$ and $f_k(H) =0$, where $\{B_k\}=\{I_{x},I_{y},I_{z},S_{x}I_{x},S_{x}I_{y},S_{x}I_{z},S_{y}I_{x},S_{y}I_{y},S_{y}I_{z},S_{z},S_{z}I_{x},S_{z}I_{y},S_{z}I_{z}\}$.
The target evolution operator is a controlled-U gate which flips the electron spin qubit iff the nuclear spin is in state $|m_I=0\rangle$. The form of this gate is shown in equation 2 in the main text.
This gate is a non-trivial two-qubit gate \cite{JAJones_S}, which can convert a product state to an entangled state.

\begin{figure}[!h]\centering
\includegraphics[width=0.7\columnwidth]{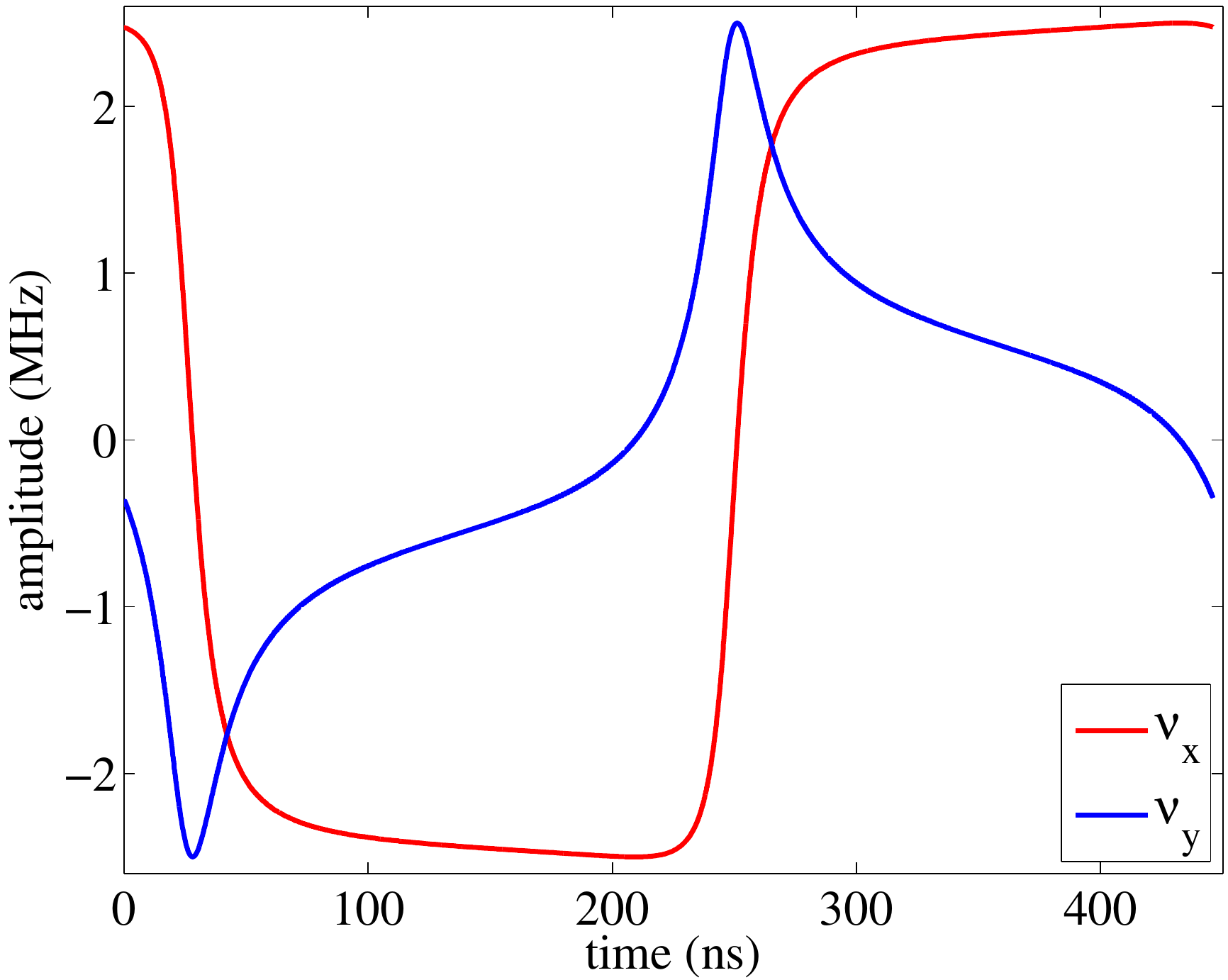}
\caption{\label{pulse sequence} \textbf{Pulse sequence for non-trivial two-qubit gate, $U_c$, via TOC in NV system.} The amplitude of the control pulse is $\nu_{1}=2.5$~MHz. The x- and y-components of the control amplitude are denoted by $\nu_x$ and $\nu_y$.}
%The component of amplitude in $\hat x$ and $\hat y$ axes are denoted by $\nu_x$ and $\nu_y$.}
\end{figure}

The time-optimal control Hamiltonian $H_c(t)$ can be numerically obtained by steps mentioned above, with $\mathcal P  =$span$\{ S_x, S_y\}$, $\mathcal Q =$span$\{B_k\}$ and $q = 1000$.
%The amplitude of time-optimal pulse is time independent with $\nu_1 = 2.5~$MHz.
Figure \ref{pulse sequence} shows the time-optimal pulse sequence, where $\nu_x={\nu_1}\cos\phi(t)$ and $\nu_y={\nu_1}\sin\phi(t)$ are x- and y-components of the control amplitude.
Although the time-optimal sequence is derived for a closed system, high fidelity is expected when applying it to realistic systems.
For example, if considering dephasing effect and imperfection of control field \cite{NatCommun_Du_S}, an fidelity of 0.9933 is estimated.
The high fidelity benefits from the least possible accumulation of errors during the shortest possible time, which is 446.1~ns.

We compare the time duration of controlled-U gate with TOC and that with a conventional method of applying a selective pulse\cite{PRA_Dorai_S, JMR_Mahesh_S}.
The frequency of the pulse is matched to the energy difference between $|m_S=0, m_I=0\rangle$ and $|m_S=-1, m_I=0\rangle$.
By choosing an appropriate control strength, the controlled-U gate can be realized with conditional rotation of the electron spin, i.e. a rotation of $(2k_1+1)\pi$ for $|m_I=0\rangle$ and $2k_2\pi$ for $|m_I=1\rangle$, where $k_1$ and $k_2$ are integers.
%This controlled-U gate is accomplished though the rotating electron spin of $(2k_1+1)\pi$ with the nuclear spin in $|m_I=0\rangle$ and a rotation of $2k_2\pi$ with the nuclear spin in $|m_I=1\rangle$, where $k_1$ and $k_2$ are integers.
%In this scenario, the electron spin must undergo a rotation of $(2k_1+1)\pi$ if the nuclear spin is in $|m_I=0\rangle$, and a rotation of $2k_2\pi$ with the nuclear spin in $|m_I=1\rangle$, where $k_1$ and $k_2$ are integers.
%This can be accomplished by applying a pulse on resonance with the transition $|m_S=0, m_I=0\rangle$ to $|m_S=-1, m_I=0\rangle$ and carefully selecting the control strength that satisfies $2\pi\omega_1T=(2k_1+1)\pi$ and $2{\pi}\sqrt{\omega_1^2+A^2}T=2k_2\pi$, where $T$ is the total duration of the pulse.
%We find that $\omega_1=2.5$~MHz approximately satisfies the condition with $k_1=1$ and $k_2=2$.
%With this method, a unitary gate of fidelity 0.994 with the controlled-U gate can be generated.
With $\nu_1=2.5$~MHz, the controlled-U gate can be realized by $k_1 = 1$ and $k_2 = 2$.
The time duration with this method is 612.4~ns, which is more than 160~ns longer than that with TOC.
%We generate a unitary gate of fidelity 0.994 with the controlled-U gate by choosing $\omega_1=2.5$~MHz.
%The pulse duration of this case is 612.4~ns, which is 166.3~ns longer than the duration of our time-optimal pulse sequence.
%The amplitude component in $\hat x$ and $\hat y$ axes are $\nu_x={\nu_1}\cos\phi(t)$ and $\nu_y={\nu_1}\sin\phi(t)$ which are shown in FIG.~\ref{pulse sequence}.
%The performance of the gate is characterized by the average fidelity which is estimated to be $F_a=0.9933$ , with dephasing effect and imperfect control field taken into consideration.

\iffalse
\begin{figure}\centering
\includegraphics[width=0.7\columnwidth]{TwoQubitEvoluEnergyDiag.pdf}
\caption{\label{energy level diagram} \textbf{Energy level diagram and state populations before and after the two-qubit time-optimal control $U_c$.} The quantum state of the two-qubit system
is denoted as $\mathrm{|m_S, m_I\rangle}$. The green circles denote the occupation of the energy level. The arrows between the energy levels indicate the sate evolution under the control-U gate.}
\end{figure}
\fi

\section{\uppercase\expandafter{\romannumeral3}. Quantum process tomography}
%\subsection{Quantum process tomography}
\textrm{\\}
We use standard quantum process tomography \cite{Nielsen_S} to evaluate the experimentally realized quantum gates. An unknown process $\mathcal{E}$ acting on the initial state $\rho$ and generating the final state ${\mathcal{E}}(\rho)$ can be described as
\begin{equation}
\mathcal{E}(\rho)=\sum_{m,n=1}^{d^2}\chi_{mn}A_m{\rho}A_n^{\dagger},
\end{equation}
where $A_m \in SU(d)$ represents a full set of orthogonal basis operators and $\chi_{mn}$ is the coefficient of the process matrix $\chi$ which completely describes the process $\mathcal{E}$.
%In our experiments the process matrix $\chi$ is determined in the Pauli basis $A_n=\{I, X, Y, Z\}$, where $I$ is the identity operator, and $X=\sigma_x$, $Y=\sigma_y$, and $Z=\sigma_z$ are Pauli operators.
In our experiments the process matrix $\chi$ is determined in the Pauli basis $A_n$, where $A_n=\{\sigma_i\}$ for single-qubit case and $A_n=\{\sigma_i\otimes\sigma_j\}$ for two-qubit case, $\sigma_{i(j)}\in\{I, X, Y, Z\}$, and $I$ is the identity operator, $X=\sigma_x$, $Y=\sigma_y$, and $Z=\sigma_z$ are Pauli operators.
We prepare a complete set of basis states $\rho_1\ldots\rho_{d^2}$ with microwave (MW) and radio frequency (RF) pulses. The MW pulse is applied after RF pulse to depress the decoherence of the electron spin in the preparation of two-qubit initial states. The fidelity of the quantum process is then given by the average gate fidelity \cite{PLA303_249_S},
\begin{equation}
F_{\textrm{a}}(\mathcal{E},U)=\frac{\sum_{j}\mathrm{tr}(UU_j^{\dagger}U^{\dagger}\mathcal{E}(U_j))+d^2}{d^2(d+1)},
\end{equation}
where $U$ is the theoretically ideal transformation and $U_j$ is a basis of unitary operators.

\section{\uppercase\expandafter{\romannumeral4}. Normalization of the experimental data}
%\subsection{Normalization of the experimental data}
\textrm{\\}
\begin{figure}[!h]\centering
\includegraphics[width=0.7\columnwidth]{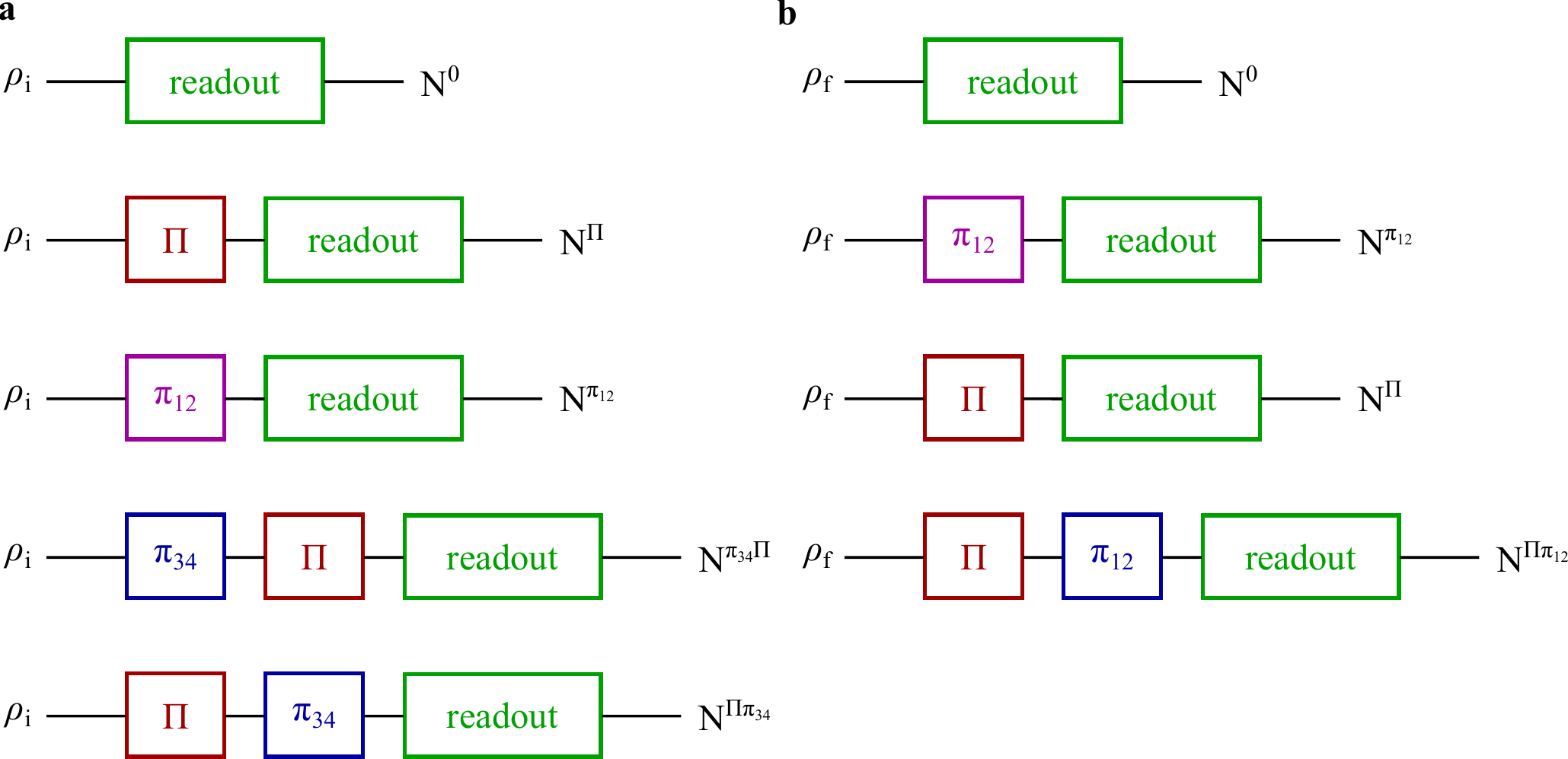}
    \caption{\textbf{Schematic normalization sequences in the two-qubit experiment.} Here $\rho_\textrm{i}$ denotes the initialized state after a laser pulse, $\rho_\textrm{f}$ denotes the final state after applying control sequence to $\rho_\textrm{i}$. $\pi_{12}$ ($\pi_{34}$) is a selective $\pi$ pulse on transition between states $|m_S=0, m_I=1\rangle$ and $|m_S=0, m_I=0\rangle$ (transition between states $|m_S=-1, m_I=1\rangle$ and $|m_S=-1, m_I=0\rangle$). $\Pi$ indicates a non-selective $\Pi$ pulse on the electron spin-qubit. The length of non-selective $\Pi$ pulse is $20~$ns.  $N^x$ indicates the detected photoluminescence intensity after applying the pulse sequence $x$.
 }
    \label{TwoQubitNorm}
\end{figure}

In the single-qubit experiment,the normalization is carried out by performing a nutation experiment\cite{PRL112_050503_S}. The normalized data corresponds to the population of $\mathrm{|0\rangle}$ for the final state.

In the two-qubit experiment, the population of $\mathrm{|0,1\rangle}$, $\mathrm{|0,0\rangle}$, $\mathrm{|-1,1\rangle}$ and $\mathrm{|-1,0\rangle}$ ($P_{0,1}$, $P_{0,0}$, $P_{-1,1}$and $P_{-1,0}$) for the final state is obtained by normalization. According to Ref.~\onlinecite{Nat484_82_S}, each occupied energy level contributes to the measured photoluminescence intensity $(I_{PL})$ with a different PL rate and these different PL rates are measured and used to determine the population of the levels with several  sequences. Here we describe this set of measurements. For brevity of notation in the following equations, we relabel the states $\mathrm{|0,1\rangle}$, $\mathrm{|0,0\rangle}$, $\mathrm{|-1,1\rangle}$, $\mathrm{|-1,0\rangle}$ as 1,2,3,4 respectively.

The number of photons we detect upon PL readout of the initialized state is
\begin{equation}
N^0=enN_1+e(1-n)N_2+(1-e)N_3+(1-e)(1-n)N_4,
\end{equation}
where $N_i$ is the number of detected photons if all of the population within the two-qubit subspace occupies level $i$, and $e (n)$ is the fraction of the population within the two-qubit subspace in the levels 1 and 2 (1 and 3). We determine $N_i$ and $e$, while $n$ is approximated to 1, by applying a set of pulse sequences to the initial state and measuring the PL as shown in Fig.~\ref{TwoQubitNorm}. This yields
\begin{equation}
\begin{bmatrix}
e & 0 & 1-e & 0 \\
1-e & 0 & e & 0 \\
0 & e & 1-e & 0 \\
0 & 1-e & e & 0 \\
1-e & 0 & 0 & e \\
\end{bmatrix}
\begin{bmatrix}
 N_1 \\ N_2 \\ N_3 \\ N_4 \\
 \end{bmatrix}
= \begin{bmatrix}
N^0 \\ N^{\Pi} \\ N^{\pi_{12}} \\ N^{\pi_{34}\Pi} \\ N^{\Pi\pi_{34}} \\ \end{bmatrix},
\end{equation}
where $N^x$ indicates the detected PL after applying the pulse sequence $x$, $\pi_{ij}$ indicates a selective $\pi$ pulse on transition $i \Leftrightarrow j$, and $\Pi$ indicates a non-selective $\Pi$ pulse on the electronic transition (which flips both transitions $\mathrm{1} \Leftrightarrow \mathrm{3}$ and $\mathrm{2} \Leftrightarrow \mathrm{4}$).
Knowing the $N_i$ we can determine the occupation probabilities of an arbitrary state of the system. The PL of an arbitrary state with level occupation probabilities $p_i$ is
\begin{equation}
N^0=p_1N_1+p_2N_2+p_3N_3+p_4N_4.
\end{equation}
By flipping populations within the two-qubit subspace and measuring the resulting PL we can calculate the $p_i$ from
\begin{equation}
\begin{bmatrix}
N_1 & N_2 & N_3 & N_4 \\
N_2 & N_1 & N_3 & N_4 \\
N_3 & N_4 & N_1 & N_2 \\
N_4 & N_3 & N_1 & N_2 \\
\end{bmatrix}\begin{bmatrix} p_1 \\ p_2 \\ p_3 \\ p_4 \\ \end{bmatrix} = \begin{bmatrix} N^0 \\ N^{\pi_{12}} \\ N^{\Pi} \\ N^{\Pi\pi_{12}} \\ \end{bmatrix}.
\end{equation}
The set of pulse sequences is shown in Fig.~\ref{TwoQubitNorm}.

\end{document}